\documentclass[fleqn,usenatbib]{mnras}
\usepackage{newtxtext,newtxmath}
\usepackage[T1]{fontenc}

\DeclareRobustCommand{\VAN}[3]{#2}
\let\VANthebibliography\thebibliography
\def\thebibliography{\DeclareRobustCommand{\VAN}[3]{##3}\VANthebibliography}

\usepackage{graphicx}	% Including figure files
\usepackage{amsmath}	% Advanced maths commands
\usepackage[left]{lineno}
\usepackage{pdflscape}
\usepackage{longtable}
\usepackage{threeparttable}

\newcommand{\Msun}{\mathrm{M_\odot}}
\newcommand{\EW}{\mathrm{EW_0}}

\title[Low-luminosity quasar environments at $z\sim6$]{Subaru High-$z$ Exploration of Low-Luminosity Quasars (SHELLQs). XXV. Large-scale environments of low-luminosity quasars at $z\sim6$ traced by Ly$\alpha$ emitters}

\author[J. Arita et al.]{
Junya Arita,$^{1}$\thanks{E-mail: jarita@astron.s.u-tokyo.ac.jp}
Nobunari Kashikawa,$^{1,2}$
Yoshiki Matsuoka,$^{3}$
Masafusa Onoue,$^{4,5}$
Michael A. Strauss,$^{6}$
\newauthor
Kentaro Koretomo,$^{1}$
Yoshihiro Takeda,$^{1}$
Ryo Emori,$^{1}$
Wanqiu He,$^{7}$
Hiroki Hoshi,$^{1}$
Masatoshi Imanishi,$^{7,8}$
\newauthor
Rikako Ishimoto,$^{1}$
Kei Ito,$^{9, 10}$
Kazushi Iwasawa,$^{11, 12}$
Satoshi Kikuta,$^{1}$
Yongming Liang,$^{7, 13}$
Camryn L. Phillips,$^{6}$
\newauthor
Shunta Shimizu,$^{1}$
John D. Silverman,$^{1, 5, 14, 15}$
Yoshiki Toba,$^{3,16,17}$
Takehiro Yoshioka$^{1}$
\\
$^{1}$Department of Astronomy, School of Science, The University of Tokyo, 7-3-1, Hongo, Bunkyo, Tokyo 113-0033, Japan \\
$^{2}$Center for the Early Universe, The University of Tokyo, 7-3-1 Hongo, Bunkyo, Tokyo, 113-0033, Japan\\
$^{3}$Research Center for Space and Cosmic Evolution, Ehime University, 2-5 Bunkyo-cho, Matsuyama, Ehime 790-8577, Japan\\
$^{4}$Waseda Institute for Advanced Study (WIAS), Waseda University, 1-21-1, Nishi-Waseda, Shinjuku, Tokyo 169-0051, Japan\\
$^{5}$Kavli Institute for the Physics and Mathematics of the Universe (Kavli IPMU, WPI), UTIAS, Tokyo Institutes for Advanced Study, University of Tokyo, Chiba, \\277-8583, Japan\\
$^{6}$Department of Astrophysical Sciences, Princeton University, 4 Ivy Lane, Princeton, NJ 08544, USA\\
$^{7}$National Astronomical Observatory of Japan, Mitaka, Tokyo 181-8588, Japan\\
$^{8}$Department of Astronomical Science, Graduate University for Advanced Studies (SOKENDAI), Mitaka, Tokyo 181-8588, Japan\\
$^{9}$Cosmic Dawn Center (DAWN), Copenhagen, Denmark\\
$^{10}$DTU Space, Technical University of Denmark, Elektrovej 327, DK2800 Kgs. Lyngby, Denmark\\
$^{11}$Institut de Ci\`{e}ncies del Cosmos (ICCUB), Universitat de Barcelona (IEEC-UB), Mart\'{i} i Franqu\`{e}s, 1, 08028 Barcelona, Spain\\
$^{12}$ICREA, Pg Llu\'{i}s Companys 23, 08010 Barcelona, Spain\\
$^{13}$Institute for Cosmic Ray Research, The University of Tokyo, 5-1-5 Kashiwanoha, Kashiwa, Chiba 277-8582, Japan\\
$^{14}$Center for Data-Driven Discovery, Kavli IPMU (WPI), UTIAS, The University of Tokyo, Kashiwa, Chiba 277-8583, Japan\\
$^{15}$Center for Astrophysical Sciences, Department of Physics \& Astronomy, Johns Hopkins University, Baltimore, MD 21218, USA\\
$^{16}$Department of Physical Sciences, Ritsumeikan University, 1-1-1 Noji-higashi, Kusatsu, Shiga 525-8577, Japan\\
$^{17}$Academia Sinica Institute of Astronomy and Astrophysics, 11F of Astronomy-Mathematics Building, AS/NTU, No.1, Section 4, Roosevelt Road, Taipei 10617, Taiwan
}

\date{Accepted XXX. Received YYY; in original form ZZZ}

\pubyear{\the\year{}}

\begin{document}
\label{firstpage}
\pagerange{\pageref{firstpage}--\pageref{lastpage}}
\maketitle

% show line numbers (activated by \usepackage{linno})
% \linenumbers

\begin{abstract}
High-$z$ quasars are believed to reside in massive dark matter haloes (DMHs), suggesting that they reside in galaxy overdense regions.
However, previous observations have shown a range of environments around them.
These fields have been limited to luminous quasars ($M_{1450}\lesssim-25$), for which photoevaporation may hinder galaxy formation in their vicinity.
Here, we present Subaru/Hyper-Suprime Cam observations of the environments of four low-luminosity quasars ($-24<M_{1450}<-22$) at $z\sim6.18$, which are expected to have a smaller photoevaporation effect.
We detect Lyman $\alpha$ emitters (LAEs) with narrowband NB872 imaging, and measure the local LAE overdensity.
One quasar (J0844$-$0132) resides in an overdense region ($\delta_\mathrm{LAE}=1.97\pm0.40$), whereas the other three fields are consistent with no overdensity.
These results hold over the proximity zone of each quasar, suggesting that the diverse environment around quasars is independent of photoevaporation.
We find no significant correlation between the LAE overdensities and the characteristics of host galaxies and supermassive black holes.
Our quasars have host stellar mass measurements from \textit{JWST}, allowing us to compare them with the LAE overdensity around galaxies without quasar activity with comparable stellar masses.
We find that the LAE overdensity in the J0844$-$0132 field is stronger than that of galaxies with similar stellar mass at $z\sim6$.
\end{abstract}


%%%%%%%%%%%%%%%%%%%%%%%%%%%%%%%%%%%%%%%%%%%%%%%%%%

%%%%%%%%%%%%%%%%% BODY OF PAPER %%%%%%%%%%%%%%%%%%

\section{Introduction} \label{sec:intro}
The emergence of supermassive black holes (SMBHs) within 1 Gyr after the Big Bang, as revealed by recent observations of quasars at $z\gtrsim6$ (e.g. \citealp{Mortlock2011, Banados2018, SHELLQs7, Yang2020, Wang2021} and \citealp{Fan2023_review} for review), is one of the most intriguing mysteries in the early Universe.
To explain their rapid growth, it is suggested that high-$z$ quasars reside in massive dark matter haloes (DMHs), which hold a reservoir of sufficient gas for mass growth \citep{Volonteri2006}.
This picture is supported by a variety of cosmological simulations (e.g. \citealp{Springel2005, Costa2014}).
However, observational evidence for this has been lacking due to the very low number density of high-$z$ luminous quasars that have been identified in previous large surveys, which hinders us from measuring their clustering, and thus constraining their DMH mass (e.g. \citealp{Martini2001}).
Measuring the clustering signal of high-$z$ quasars with neighbouring galaxies with high statistical significance requires a wide and deep survey of the quasars themselves and their surrounding galaxies; however, such surveys have typically been very expensive.

Recently, \citet{Arita2023} measured the auto-correlation function of low-luminosity ($M_{1450}\lesssim-21$) quasars at $z\sim6$.
They found that their typical DMH mass is high, $\sim10^{12.85}\,\Msun$, at the massive end of the DMH mass function at $z\sim6$, suggesting that quasars are born in rare high-density peaks.
The result is supported by the cross-correlation of bright quasars at $z\sim6$ with surrounding [\ion{O}{iii}] emitters observed by \textit{JWST}/NIRCam slitless spectroscopy in Emission-line galaxies and Intergalactic Gas in the Epoch of Reionization (EIGER) program \citep{Eilers2024} and A SPectroscopic survey of biased halos In the Reionization Era (ASPIRE) program \citep{Huang2026}.
\citet{Eilers2024} and \citet{Huang2026} derived a minimum DMH mass of $10^{12.43}\,\Msun$ and $10^{12.13}\,\Msun$ using four bright quasars with $92$ [\ion{O}{iii}] emitters and 25 bright quasars with 487 [\ion{O}{iii}] emitters, respectively.
These studies also show that the bias parameter of quasars tends to increase as the redshift increases, which suggests that higher-$z$ quasars reside in rarer high density peaks than lower-$z$ quasars.
In fact, previous statistical studies on the large-scale environments of quasars at $z<4$ show that many of them reside in non-overdense regions \citep{Uchiyama2018, Uchiyama2020, Suzuki2024, Shibata2025}.
Contrary to the large-scale environment of the lower-$z$ quasars, that of $z\sim6$ quasars is highly biased, and cosmological simulations suggest that quasars reside in galaxy overdensities extending up to $\sim 10$ pMpc \citep{Overzier2009}.
Understanding the environment of early quasars is essential to reveal the amount of gas fuelling SMBHs and the mechanism of gas accretion onto them (e.g. \citealp{Granato2004}), and also has significant implications for the onset of the mass-related coevolution between galaxies and SMBHs.

However, observational evidence of galaxy clustering around quasars at $z\sim6$ is far from conclusive, with some studies reporting significant overdensities (e.g. \citealp{Zheng2006, Utsumi2010, Morselli2014, Mignoli2020, Wang2024, Lambert2024}), while others find normal or even underdense environments (e.g. \citealp{Banados2013, Simpson2014, Mazzucchelli2017, Kikuta2017, Goto2017}).  
Many of these studies used the distribution of Lyman-break galaxies (LBGs) or Ly$\alpha$ emitters (LAEs), which are selected based on photometry.
They had very different fields of view (FoVs) in these studies, from only a few arcmin$^2$ to $>1$ deg$^2$, but most of them are smaller than the expected galaxy overdensity scale.
Recently, some studies performed \textit{JWST} observations to investigate the quasar environments at $z\gtrsim6$.
\citet{Kashino2023} and \citet{Wang2023} found a clear overdensity of [\ion{O}{iii}] emitters around the luminous quasars J0100$+$2802 and J0305$-$3150 using the NIRCam slitless spectroscopy of $22.1$ and $\sim11$ arcmin$^2$ fields, respectively.
\citet{Champagne2025} confirmed a 10 cMpc overdensity structure of [\ion{O}{iii}] emitters in the J0305$-$3150 field using the \textit{JWST}/NIRCam imaging and slitless spectroscopy, covering a 35 arcmin$^2$ field.
\citet{Wang2026} extended these studies to 28 quasars in ASPIRE to report that seven quasars out of them show strong overdensities of [\ion{O}{iii}] emitters, but the individual effective areas are limited to a single \textit{JWST} FoV.
While these \textit{JWST}/NIRCam observations can immediately confirm their redshift spectroscopically, the FoV is still smaller than the expected overdensity scale.
The lack of clear evidence for galaxy overdensities around $z\sim6$ quasars is a real challenge to our current paradigm of early black hole growth.

One possible reason why galaxy overdensities around $z\sim6$ quasars are not always seen is feedback by the quasars themselves on the surrounding gas.
Photoionization heating by intense UV radiation from quasars could heat the collapsed gas in a galactic halo and inhibit gas cooling \citep{Barkana1999}, suppressing star formation and delaying galaxy growth against galaxy self-shielding. 
Here, we refer to the intense photoionization heating which could hinder or delay star formation in surrounding galaxies as quasar photoevaporation.
Some observations have found evidence of the effect at lower-$z$ \citep{Kashikawa2007, Banados2018, Uchiyama2019}.
The effect of the photoevaporation is stronger in more luminous quasars.
An ionized region around quasars generated by their own radiation is called the proximity zone \citep{Cen2000}, and its size $R_\mathrm{p}$, which depends on quasar luminosity, can be recognized as the scale over which photoevaporation works (e.g. \citealp{Suzuki2025}).
Most previous studies of galaxy overdensities around quasars have focused on scales within $R_\mathrm{p}$, which is approximately $4.71$ pMpc at $M_{1450}=-27$ at $z\sim6$ and proportional to $R_\mathrm{p}\propto10^{-0.12M_{1450}}$
\citep{Eilers2017}, due to the limited FoV of the instruments (e.g. \citealp{Willott2005, Zheng2006, Kim2009, Banados2013, Mazzucchelli2017, Mignoli2020}).
These studies investigate highly limited areas in quasar fields, probing only dozens of arcmin$^2$ around bright quasars, where the photoevaporation effect can suppress star formation, decreasing the number of galaxies observed despite the presence of an underlying massive DMH.
Quasar photoevaporation, in principle, is a serious consideration for understanding the intrinsic quasar DMH mass and environment in the early universe.
The intrinsic galaxy clustering strength can best be measured in regions outside of $R_\mathrm{p}$ and inside the expected cross-correlation scale of galaxies and quasars ($>24\mathchar`-$ cMpc; \citealp{Mignoli2020}), which requires investigating the lower-luminosity quasar fields where $R_\mathrm{p}$ is smaller.

To minimize the photoevaporation effect, we investigate the fields of low-luminosity quasars whose expected proximity zone sizes are quite smaller than the expected overdensity scale.
We select the low-luminosity quasars discovered by the Subaru High-$z$ Exploration of Low-Luminosity Quasars (SHELLQs; \citealp{SHELLQs1}) project.
Their absolute magnitudes lie in the range of $-24< M_{1450}<-22$, which is $\gtrsim2$ mag fainter than those of bright quasars investigated in previous studies.
Many studies have found that the DMH mass and thus presumably the corresponding galaxy overdensity are independent of $M_{1450}$ at low-$z$ \citep{Croom2005, Shen2009} and $z\sim6$ \citep{Arita2023}, suggesting that targeting low-luminosity quasars does not result in examining a biased environment.
Moreover, the targeted quasars have been observed photometrically and spectroscopically by \textit{JWST} (GO 1967, PI: M. Onoue), which provides important host galaxy properties such as stellar mass ($M_*$) \citep{Ding2023, Ding2025} and star formation rate (SFR) \citep{Phillips2025} and SMBH properties of SMBH mass and Eddington ratio (Onoue et al. in prep.).
We use Hyper Suprime-Cam (HSC; \citealp{Miyazaki2018, Komiyama2018, Kawanomoto2018, Furusawa2018}) on the Subaru Telescope, whose FoV corresponds to roughly 220 cMpc at $z\sim6$.
HSC has a series of narrowband filters, including NB872 ($\lambda_c=8721$~\AA, $\mathrm{FWHM}=67$~\AA), which corresponds to Ly$\alpha$ emission at $z=6.174\pm0.028$.
In this paper, we, for the first time, present the low-luminosity quasar environments at $z\sim6$ traced by LAEs, generally young low-mass galaxies, and the relation between LAE overdensity and the quasar properties.

This paper is structured as follows.
The details of the observations and LAE selection are described in Sec. \ref{sec:data}, and their results are shown in Sec. \ref{sec:results}.
In Sec. \ref{sec:discussion}, we discuss the environment of low-luminosity quasars by comparing that of bright quasars investigated in the previous studies and the relation between the quasar environment and quasar properties.
Finally, we summarize this paper in Sec. \ref{sec:summary}.
We adopt a flat $\Lambda$ cold dark matter cosmology with $h=0.7, \Omega_m=0.3$ throughout this paper.
All magnitudes in this paper are presented in the AB system \citep{Oke1983}.
The notations of cMpc and pMpc are used to indicate the comoving and physical scales, respectively.

\section{Observations} \label{sec:data}
We select four low-luminosity ($M_{1450}>-24$) quasars in the SHELLQs project: HSC J084408.61$-$013216.5, HSC J091833.17$+$013923.4, HSC J142517.72$-$001540.8, and HSC J151248.71$+$442217.5 (hereafter J0844$-$0132, J918$+$0139, J1425$-$0015, and J1512$+$4422), to investigate their environment using the surrounding LAEs.
These quasars have accurate systemic redshifts of $z\sim6.18$, so surrounding Ly$\alpha$-emitting galaxies will appear in NB872.
The quasars were first discovered in the HSC Subaru Strategic Program (HSC-SSP) field \citep{Aihara2018} and have been spectroscopically confirmed with the Faint Object Camera and Spectrograph (FOCAS; \citealp{Kashikawa2002}) on the Subaru telescope.
Accurate redshift measurements of the quasars based on [\ion{O}{III}]$\lambda5007$ emission line are available from \textit{JWST} NIRSpec spectra (Onoue et al. in prep.).
Their properties are summarized in Table \ref{tab:qso_properties}.

\begin{table*}
    \caption{Overview of the physical properties of the target quasars and the effective area of our HSC observations for the individual objects.
    All of the quasar redshifts are the systemic redshifts determined by the [\ion{O}{iii}]$\lambda5007$ line (Onoue et al. in prep.).}
    \label{tab:qso_properties}
    \centering
    \begin{tabular}{lcccccc} \hline
        Field & Effective area & R.A. & Decl. & $M_{1450}$ & $z_\mathrm{qso}$ & Discovery paper \\
         & (arcmin$^2$) & (J2000) & (J2000) & (mag) & & \\ \hline
        J0844$-$0132 & 5173 & 08:44:08.61 & $-$01:32:16.5 & $-23.97$ & $6.1830\pm0.0000 
$ & \citet{SHELLQs4} \\
        J0918$+$0139 & 4837 & 09:18:33.17 & +01:39:23.4 & $-23.71$ & $6.1786\pm0.0010 
$ & \citet{SHELLQs4} \\
        J1425$-$0015 & 5092 & 14:25:17.72 & $-$00:15:40.8 & $-23.44$ & $6.1780\pm0.0005 
$ & \citet{SHELLQs2} \\
        J1512$+$4422 & 5137 & 15:12:48.71 & +44:22:17.5 & $-22.07$ & $6.1806\pm0.0006 
$ & \citet{SHELLQs10} \\
        \hline
    \end{tabular}
\end{table*}

\subsection{Hyper Suprime-Cam imaging} \label{subsec:hsc_img}
\begin{figure}
    \centering
    \includegraphics[width=\linewidth]{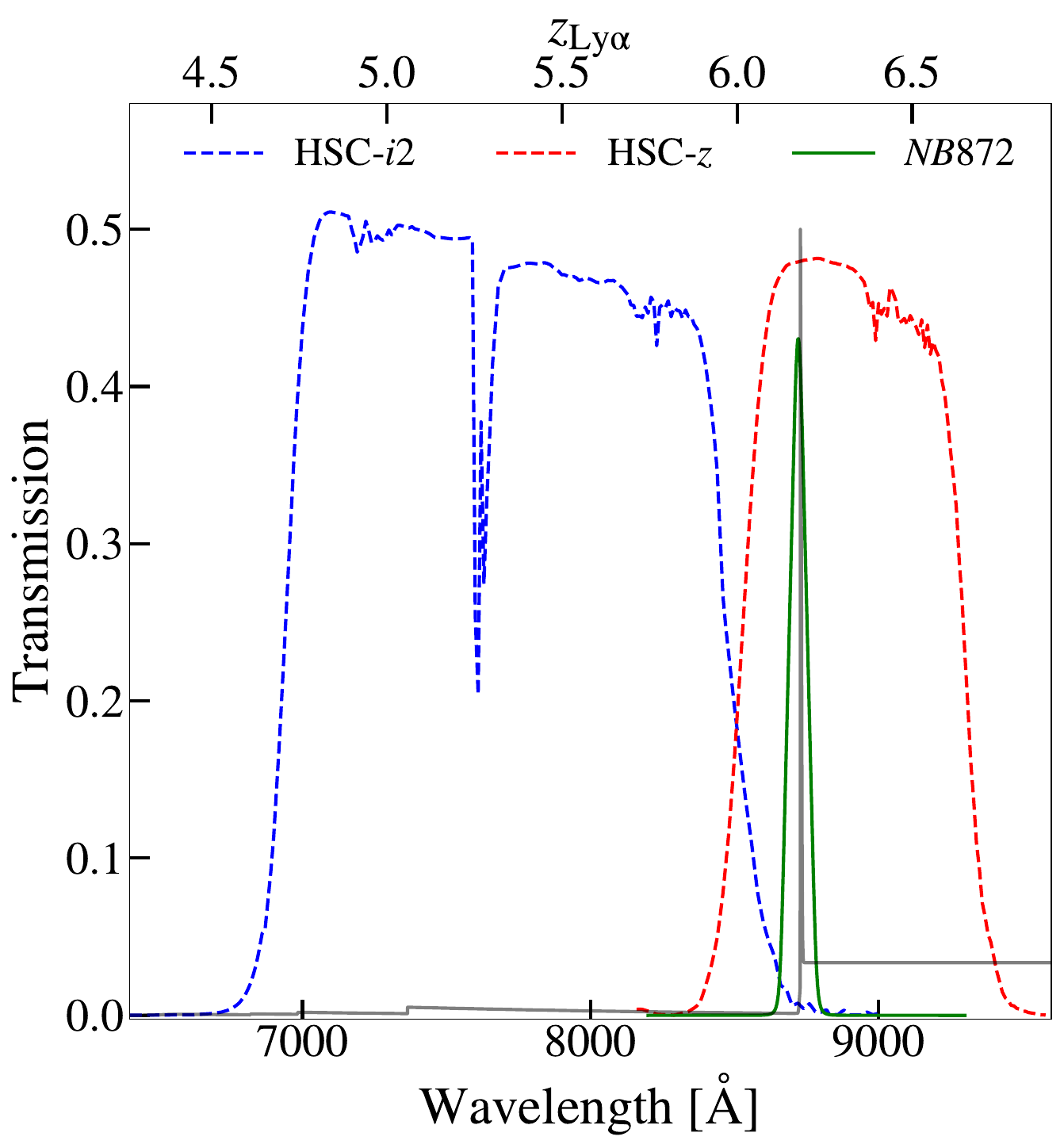}
    \caption{Filter transmission curves in this study.
    The red and blue dashed lines represent the transmission curves of HSC-$i2$ and HSC-$z$, respectively.
    The green solid line is the transmission curve of NB872.
    These transmission curves include the atmospheric absorption.
    The grey line shows a model spectrum of an LAE at $z=6.18$ generated by the procedure in Sec. \ref{subsec:lae_selection} with $\alpha=-2$ and $\EW=15$~\AA.
    The upper $x$-axis displays the redshift of Ly$\alpha$.
    }
    \label{fig:filter_trans}
\end{figure}

The four quasar fields were included in the HSC-SSP Wide layer, so five broad bands ($grizy$) of imaging data are available.
However, the exposure time of these observations is not sufficient ($\sim1200$ s) for detecting faint galaxies at $z\sim6.2$, and the NB872 imaging was not performed.
Therefore, we performed additional Subaru/HSC observations centred on the quasars with two broadband filters (HSC-$i2$ and HSC-$z$) and NB872 from December 2024 to July 2025 in queue mode.
The filter transmission curves are shown in Fig. \ref{fig:filter_trans}.
During the HSC observations, we performed dithering with a pre-defined 5-point dither pattern\footnote{\url{https://www.naoj.org/Observing/queue/pidoc/proposals.html\#dithering}} and recommended dithering parameters ($120\arcsec$)
for both RA and DEC directions to fill the gaps between the CCDs.
Individual exposure times for the broad bands and NB872 are $\sim150$ s and $900$ s, respectively.
Table \ref{tab:obs_info} gives the breakdown of total exposure time and seeing in the observations.
The pixel scale of HSC is $0\farcs168$, and the spatial resolution is shown in Table \ref{tab:obs_info}.
In the data reduction, we combined the raw broadband data from HSC-SSP with that of our observation to improve the image depth.
However, we note that HSC-$i2$ data of the J0918+0139 and J1425-0015 fields are absent since the HSC-SSP used HSC-$i$ instead of HSC-$i2$
\footnote{HSC-$i$ was replaced by HSC-$i2$ on Feb. 2016.}
.
The transmission curves of HSC-$i$ and HSC-$i2$ are slightly different \citep{Kawanomoto2018}, so we did not include the data of the HSC-$i$ in HSC-SSP.
The data were processed with \texttt{hscpipe 8.5.3} \citep{Bosch2018}\footnote{\url{https://hsc.mtk.nao.ac.jp/pipedoc_e/}}, which can create calibrated images and object catalogues from raw HSC data.
In \texttt{hscpipe}, the astrometry and photometry are calibrated based on Pan-STARRS DR1 \citep{Chambers2016}.
In \texttt{hscpipe}, source detection is performed using the multiband data, and we set the filter priorities in the order of NB872, HSC-$z$, and HSC-$i2$
\footnote{The detailed algorithm of the source detection in \texttt{hscpipe} is described in \url{https://hsc.mtk.nao.ac.jp/pipedoc/pipedoc_8_e/tutorial_e/multiband_detailed.html}.}
.
In this study, we use \texttt{ext\_convolved\_ConvolvedFlux\_2\_4\_5\_instFlux} in the photometry catalogue, which is the aperture flux measured within the $1\farcs5$ diameter aperture after an aperture correction.
We use further flags to exclude objects that are affected by saturated pixels or bad pixels, as summarized in Table \ref{tab:flag_info}.

\begin{table*}
    \caption{Summary of the new observations obtained for this study.
    The seeing is the median value of the observations, and the $5\sigma$ depth is the median value of the patches.}
    \label{tab:obs_info}
    \centering
    \begin{threeparttable}
    \begin{tabular}{llcccl}\hline
        Field & Filter & Exposure time & Seeing & $m_{5\sigma}$$^\dagger$ & Observation date \\ 
         & & (hr) & (\arcsec) & (mag) & \\ \hline
        J0844$-$0132 & HSC-$i2$ & 1.11 & 0.95 & 26.27 & 2024 Nov. 26 \\
        & HSC-$z$ & 1.17 & 1.31 & 25.46 & 2024 Dec. 29, 2024 Dec. 30, 2024 Dec. 31 \\
        & NB872 & 5.25 & 0.57 & 25.08 & 2024 Dec. 21, 2024 Dec. 22, 2024 Dec. 27, 2024 Dec. 28 \\
        J0918$+$0139 & HSC-$i2$ & 1.11 & 0.87 & 26.25 & 2024 Nov. 26, 2024 Dec. 1 \\
        & HSC-$z$ & 0.92 & 1.08 & 25.55 & 2024 Dec. 29, 2024 Dec. 30 \\
        & NB872 & 5.25  & 1.14 & 25.13 & 2024 Dec. 28, 2025 Jan. 2 \\
        J1425$-$0015 & HSC-$i2$ & 1.08 & 1.23 & 26.11 & 2025 Mar. 7 \\
        & HSC-$z$ & 0$^\ddagger$ & N/A & 25.00 & N/A\\
        & NB872 & 3.75 & 0.68 & 24.75 & 2025 Apr. 23 \\
        J1512$+$4422 & HSC-$i2$ & 2.38 & 1.01 & 26.76 & 2025 Feb. 23, 2025 Mar. 7 \\
        & HSC-$z$ & 0.21 & 1.26 & 24.96 & 2025 Jul. 19 \\
        & NB872 & 7.50 & 0.90 & 25.39 & 2025 Feb. 28, 2025 Mar. 5 \\ \hline
    \end{tabular}   
    \begin{tablenotes}
        \item[$\dagger$] The $5\sigma$ depth is measured for the final coadd images that are constructed from the data of HSC-SSP and our observations.
        \item[$\ddagger$]This observation was not performed as part of our HSC queue observations, so only HSC-SSP data was used.
    \end{tablenotes}
    \end{threeparttable}
\end{table*}

\begin{table*}
\caption{Summary of the applied flags.}
    \label{tab:flag_info}
    \centering
    \begin{tabular}{lccp{8.2cm}} \hline
        Flag & Filter & Value & Comment\\ \hline
        \texttt{detect\_isPrimary} & i2, z, NB872 & True & True if source has no children and is in the inner region of a coadd patch and is in the inner region of a coadd tract and is not detected in a pseudo-filter \\
        \texttt{base\_PixelFlags\_flag\_bad} & i2, z, NB872 & False  & Bad pixel in the Source footprint \\
        \texttt{base\_PixelFlags\_flag\_interpolatedCenter} & i2, z, NB872 & False & Interpolated pixel in the Source footprint \\
        \texttt{base\_PixelFlags\_flag\_saturatedCenter} & i2, z, NB872 & False & Saturated pixel in the Source footprint\\ 
         \texttt{merge\_peak\_N872} & NB872 & True & Peak detected in NB872 \\
         \texttt{base\_Blendedness\_abs} & NB872 & $<0.2$ & Measure of how much the flux is affected by neighbours: $(1 - \mathrm{child\_instFlux}/\mathrm{parent\_instFlux})$. Operates on the absolute value of the pixels to try to obtain a "de-noised" value.\\\hline
    \end{tabular}
\end{table*}

In \texttt{hscpipe}, the magnitude zeropoints of the coadded images are set to 27.0 mag/ADU.
However, there is no NB872 data in the calibration catalogue, which can cause a non-negligible error in the magnitude zeropoint.
Therefore, we correct the magnitude zeropoint of the NB872 image assuming that the magnitude zeropoint of HSC-$z$ is adequately corrected in \texttt{hscpipe} and that the stars in the fields satisfy $\langle z-NB872\rangle=0$.
The validity of the latter assumption is confirmed by computing the $z-NB872$ colour of stars in X-shooter Spectral Library Data Release 3 \citep{Verro2022} based on their spectra.
For the magnitude zeropoint calibration, we select stars from the photometry catalogue using the seventeenth data release of the Sloan Digital Sky Survey (SDSS; \citealp{SDSS17}), which classifies stars based on their morphology.
We cross-match the SDSS stars and the sources in catalogues by \texttt{hscpipe} to extract $\sim10^4$ stars in the individual fields, which are neither saturated nor contaminated by nearby bright sources.
We measure the median colours of $z-NB872$ of the stars in the individual fields and correct the magnitude zeropoint.

\begin{figure*}
    \centering
    \includegraphics[width=\linewidth]{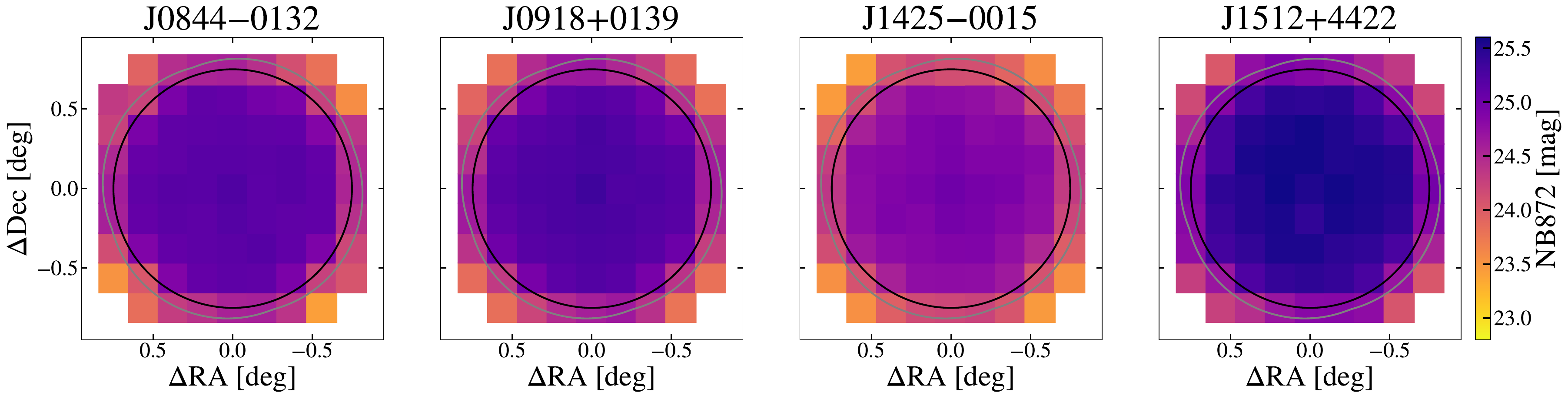}
    \caption{$5\sigma$ depth maps of the individual fields.
    The colour of each patch shows the limiting magnitudes of the NB872 image.
    The $x$-axis and $y$-axis represent the offset from the quasar coordinate.
    The black solid line displays the diameter of HSC ($45\arcmin$).
    The grey solid line shows the observed field by NB872, taking the dithering into account.
    }
    \label{fig:limiting_magnitude}
\end{figure*}

We determine mask regions around bright stars in addition to those defined in \texttt{hscpipe}.
We use Gaia Data Release 3 (DR3) \citep{Gaia2016,Gaia2023} to select bright stars ($G_\mathrm{Gaia}<18$, where $G_\mathrm{Gaia}$ is the $G$-band magnitude from Gaia DR3) in the fields.
We apply masks centred on the stars to the field, whose radii $r$ are determined by the following equation \citep{Coupon2018}:
\begin{equation}
    r\,[\arcsec] = \left\{\begin{array}{ccc}
        708.9\times\exp(-G_\mathrm{Gaia}/8.41),& \mathrm{if} & G_\mathrm{Gaia}<9,\\
        694.7\times\exp(-G_\mathrm{Gaia}/4.04),& \mathrm{if} & G_\mathrm{Gaia}\geq9.
    \end{array}
    \right.
\end{equation}
Finally, we mask regions affected by stray light, which is caused by light from bright stars or other objects reaching the detector via non-standard path in the telescope or camera.
The same masks are applied for HSC-$i2$, HSC-$z$, and NB872 images in the individual fields.
In the following analysis, we exclude the regions outside the FoV of HSC; namely, we use the inner regions of the black circle in Fig. \ref{fig:limiting_magnitude}.
After masking these regions, we measure the effective areas by scattering the random points at a number density of $100\,\mathrm{arcmin^{-2}}$ in the non-masked regions. 
The measured effective areas are 5173 arcmin$^2$, 4837 arcmin$^2$, 5092 arcmin$^2$, and 5137 arcmin$^2$ for the J0844$-$0132, J0918+0139, J1425$-$0015, and J1512+4422 fields, respectively.

We measure the $5\sigma$ depth $m_{5\sigma}$ of all filters used in this study in the individual fields, which are summarized in Table \ref{tab:obs_info}.
In \texttt{hscpipe}, the field is divided into small square regions ($\sim11\arcmin\times11\arcmin$), each of which is called a patch \citep{Bosch2018}.
We perform aperture photometry with $1\farcs5$ diameter apertures at 1,000 random positions, avoiding the masked regions for each patch.
We fit a Gaussian to the measured flux histogram and measure the $5\sigma$ depth based on its standard deviation.
The measured $5\sigma$ limiting magnitudes of the individual patches are shown in Fig. \ref{fig:limiting_magnitude}.
The rms of the limiting magnitudes of the NB872 image in the individual patches within the diameter of HSC is $\sim0.2$ mag.

\subsection{LAE selection} \label{subsec:lae_selection}

\begin{figure}
    \centering
    \includegraphics[width=\columnwidth]{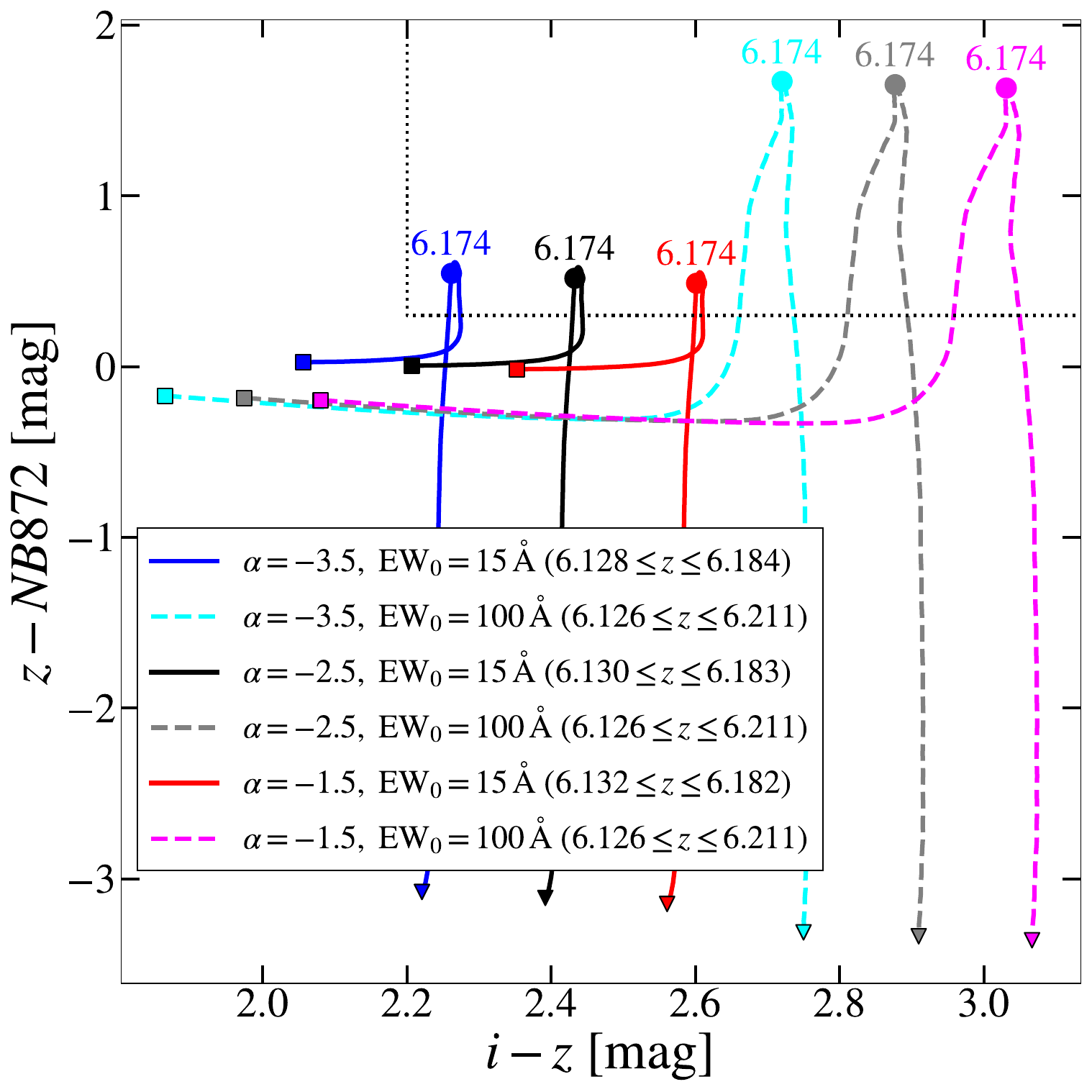}
    \caption{Colour-colour diagram of the simulated spectra.
    The solid and dashed lines represent the redshift evolution of the simulated spectra from $z=6$ (squares) to $z=6.25$ (triangles) with a step of $\Delta z = 0.001$.
    The legends display the assumed UV continuum slopes, $\EW$, and the redshift range falling within the selection box.
    The circles show the colour of LAEs at $z=6.174$, whose Ly$\alpha$ emission lies at the centre of NB872.
    The dotted lines show the colour criteria to select LAEs with $\EW>15$~\AA.
    }
    \label{fig:cc_track}
\end{figure}

We model the UV spectrum of LAEs assuming that the continuum ($f_{\lambda,\,\mathrm{cont}}$) and the Ly$\alpha$ emission ($f_{\lambda,\,\mathrm{line}}$) can be described by a power-law function and a Gaussian, respectively:
\begin{align}
    f_{\lambda,\,\mathrm{cont}}&\propto\lambda^{\alpha},\label{eq:cont}\\
    f_{\lambda,\,\mathrm{line}}&\propto\exp\left[-\frac{(\lambda-\lambda_\mathrm{Ly\alpha})^2}{2\sigma_\lambda^2}\right],    \label{eq:line}
\end{align}
where $\alpha$ is the UV spectral slope per unit wavelength, $\lambda_\mathrm{Ly\alpha}$ is the wavelength of the Ly$\alpha$ emission in the observed frame, and $\sigma_{\lambda}$ is the standard deviation of the Gaussian.
The flux ratio $f_{\lambda,\,\mathrm{line}}/f_{\lambda,\,\mathrm{cont}}$ is determined by the equivalent width ($\EW$) of the Ly$\alpha$ emission in the rest frame.
The model spectrum of the LAE is defined by summing the continuum and emission line components after the mean IGM absorption as a function of redshift from \citet{Madau1995} is applied.
We calculate the colour of an LAE with different redshifts, FWHM and thus $\sigma_\lambda$, $\alpha$, and $\EW$ to track it in colour-colour diagrams.
We find that the FWHM has little impact on the colour, hence we fix it to $250\,\mathrm{km\,s^{-1}}$, which is a reasonable value for LAEs at $z\sim6$ \citep{Prieto-Lyon2025}.
Regarding $\alpha$, we calculate the colour using $\alpha=-3.5,-2.5,-1.5$, which lie within the typical value of LAEs at $z\sim6$ \citep{Jiang2013}.
Fig. \ref{fig:cc_track} shows the track of the LAE models in the $i-z$, $z-NB872$ colour-colour diagram.
It should be noted that the FWHM of NB872 corresponds to a distance of $\sim22$ cMpc, which is larger than the expected overdensity scale ($\sim10$ cMpc).

Based on this simulation, we use the following criteria to select LAEs at $z\sim6.18$ with $\EW>15$~\AA\ (see the area outlined by black lines in Fig. \ref{fig:cc_track}):
\begin{align}
    & NB872 < NB872_{5\sigma}
    \label{eq:limiting_magnitude}\ \&\\
    &z-NB872>0.3\label{eq:colour_excess}\ \&\\
    &(i-z>2.2\ \mathrm{or}\ i>i_{3\sigma}),
    \label{eq:lyman_break}
\end{align}
where $NB872_{5\sigma}$ is the $5\sigma$ depth of the NB872 image, and $i_{3\sigma}$ represents the $3\sigma$ depth of HSC-$i2$.
If the broadband magnitudes are fainter than the $2\sigma$ depth in the field, we replace the magnitude with the $2\sigma$ depth to calculate colours.
We note that the selected LAE candidates do not change if we apply $i-z>2.4$, which corresponds to $\alpha\geq-2.5$, instead of $i-z>2.2$ in equation (\ref{eq:lyman_break}).

\begin{figure*}
    \centering
    \includegraphics[width=2\columnwidth]{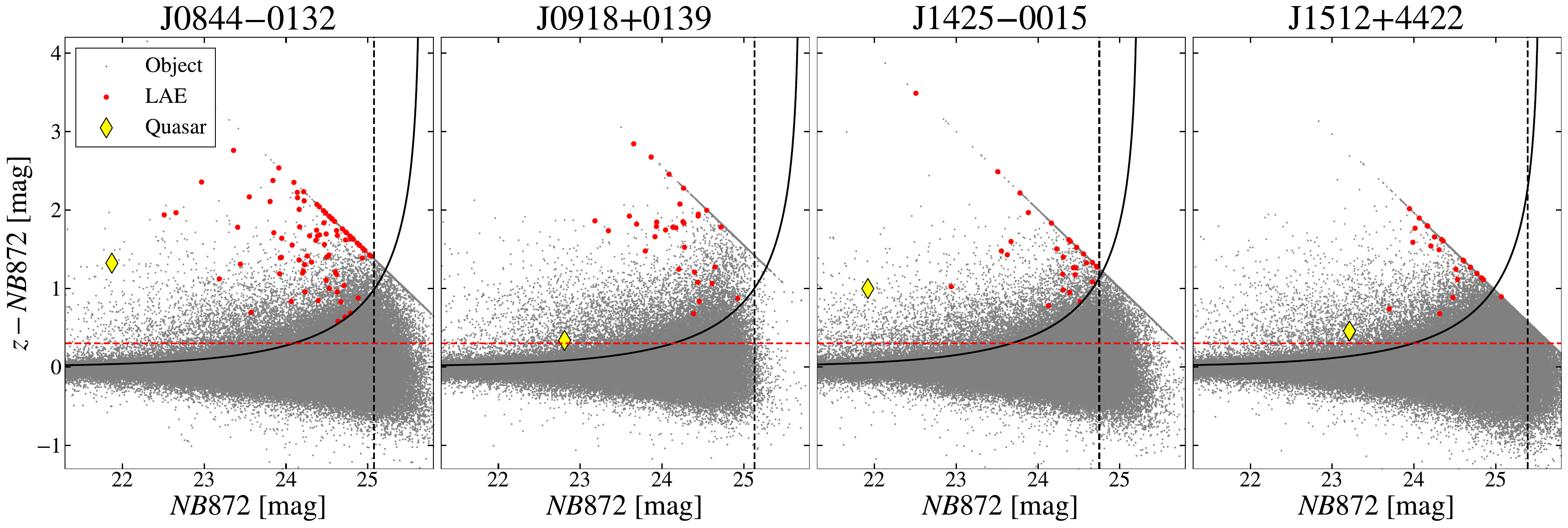}
    \caption{Colour-magnitude diagrams in the individual quasar fields.
    The grey points show the detected objects.
    The red points represent the selected LAEs, and the yellow diamonds indicate the quasars' colours.
    We replace the $z$-band magnitude with its $2\sigma$ limiting magnitude for the objects whose $z$-band magnitudes are fainter than the $2\sigma$ limiting magnitude.
    This gives the diagonal envelopes on the right of each panel.
    The red dashed lines denote the narrowband excess threshold of equation (\ref{eq:colour_excess}). 
    The black solid lines represent the narrowband excess parameter of equation (\ref{eq:nb_excess}) with the median $1\sigma$ flux errors.
    The black dashed lines are the $5\sigma$ limiting magnitudes of NB872.
    }
    \label{fig:colour_colour_diagram}
\end{figure*}

In addition to these criteria, we require that sources satisfy the following criterion:
\begin{align}
    \Sigma = \frac{1-10^{-0.4(z-NB872)}}{10^{-0.4(ZP_{NB872}-NB872)}\sqrt{\sigma_{z}^2+\sigma_{NB872}^2}}>3,
    \label{eq:nb_excess}
\end{align}
where $ZP_{NB872}$ is the magnitude zeropoint of the NB872 image, and $\sigma_{z}$ and $\sigma_{NB872}$ are the $1\sigma$ flux errors of HSC-$z$ and NB872 images, respectively.
$\Sigma$ is the narrowband excess parameter, which quantifies the significance of the flux excess in $NB872$ over $z$, and is used to ensure that the excess in the narrowband is not due to statistical fluctuations\citep{Bunker1995, Sobral2013}.
For selection completeness, defined as the recovery rate of LAEs selected by our colour selection for each specific $\EW$ range, the results are summarized in Appendix \ref{apx:selection_completeness}.
Finally, we perform visual inspection for all of the selected LAE candidates to remove artefacts.
We remove the objects that are contaminated by nearby bright objects or cosmic rays and that apparently seem like noise peaks.
In this process, 255, 409, 129, and 126 objects are removed, and 88, 31, 28, and 23 LAEs are finally selected in the J0844$-$0132, J0918$+$0139, J1425$-$0015, and J1512$+$4422 fields, respectively.
We note that we remove the central quasars in all of the fields from these LAEs.
Fig. \ref{fig:colour_colour_diagram} shows the colour-magnitude diagrams with the selected LAEs.

\begin{figure}
    \centering
    \includegraphics[width=\linewidth]{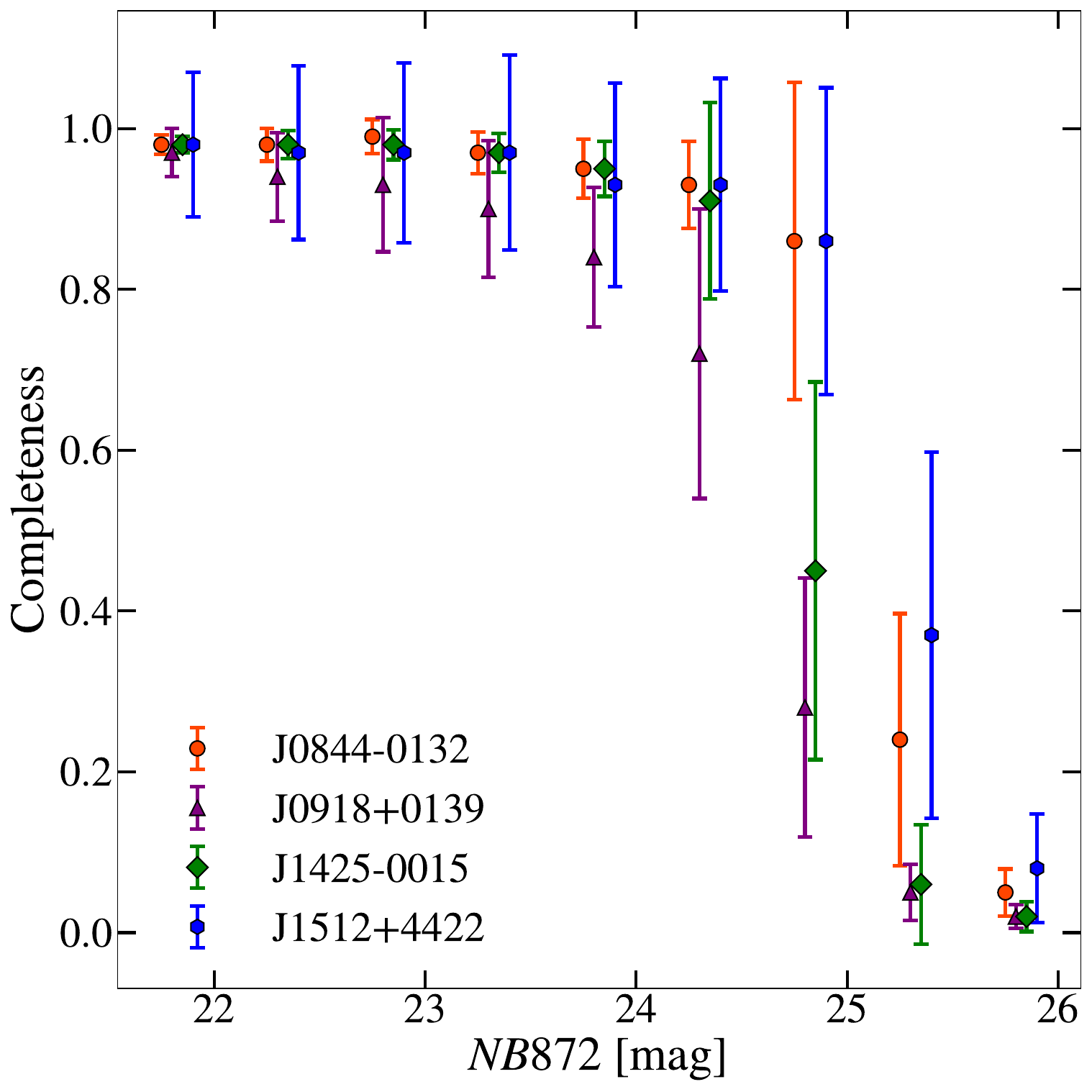}
    \caption{Detection completeness as a function of narrowband magnitudes.
    We calculate the detection completeness of patches within a FoV of HSC (black circles in Fig. \ref{fig:limiting_magnitude}).
    The median detection completeness of the patches and the scatter in the individual quasar fields are plotted.
    For visibility, the plots for individual fields are slightly offset.}
    \label{fig:detection_completeness}
\end{figure}

We estimate the detection completeness for each patch in each field using mock observations.
We create mock LAEs with \texttt{Galsim}\footnote{\url{https://github.com/GalSim-developers/GalSim}} \citep{Rowe2015} as a S\'ersic profile.
Based on the average values of LBGs at $z\sim6$ \citep{Shibuya2015}, the mock LAEs have a S\'ersic index of $n=1.5$ and a half-light radius of $r_c\sim0.9\,\mathrm{kpc}$, which corresponds to $0\farcs16$ at $z=6.18$.
The magnitudes of the mock LAEs are set in 0.5-magnitude increments from 21.75 mag to 25.75 mag.
We embed 100 mock LAEs per patch in each magnitude interval set into NB872 images after convolving with the Point Spread Function (PSF) modelled by \texttt{PSFEx}\footnote{\url{https://github.com/astromatic/psfex}} \citep{Bertin2011}.
The redshift of LAEs is fixed at $z=6.18$.
The mock injected images are processed by \texttt{hscpipe} in the same procedure as described in Sec. \ref{subsec:hsc_img}, and the completeness is calculated as the recovery rate of the mock LAEs for each patch in each field.
The median completeness as a function of source magnitude for the individual fields is shown in Fig. \ref{fig:detection_completeness}.
We note that the effect of visual inspection is not taken into account when computing the detection completeness, which may overestimate it.

\section{Results} \label{sec:results}
\subsection{Surface number density of LAEs} \label{subsec:n_lae}
\begin{figure}
    \centering
    \includegraphics[width=\columnwidth]{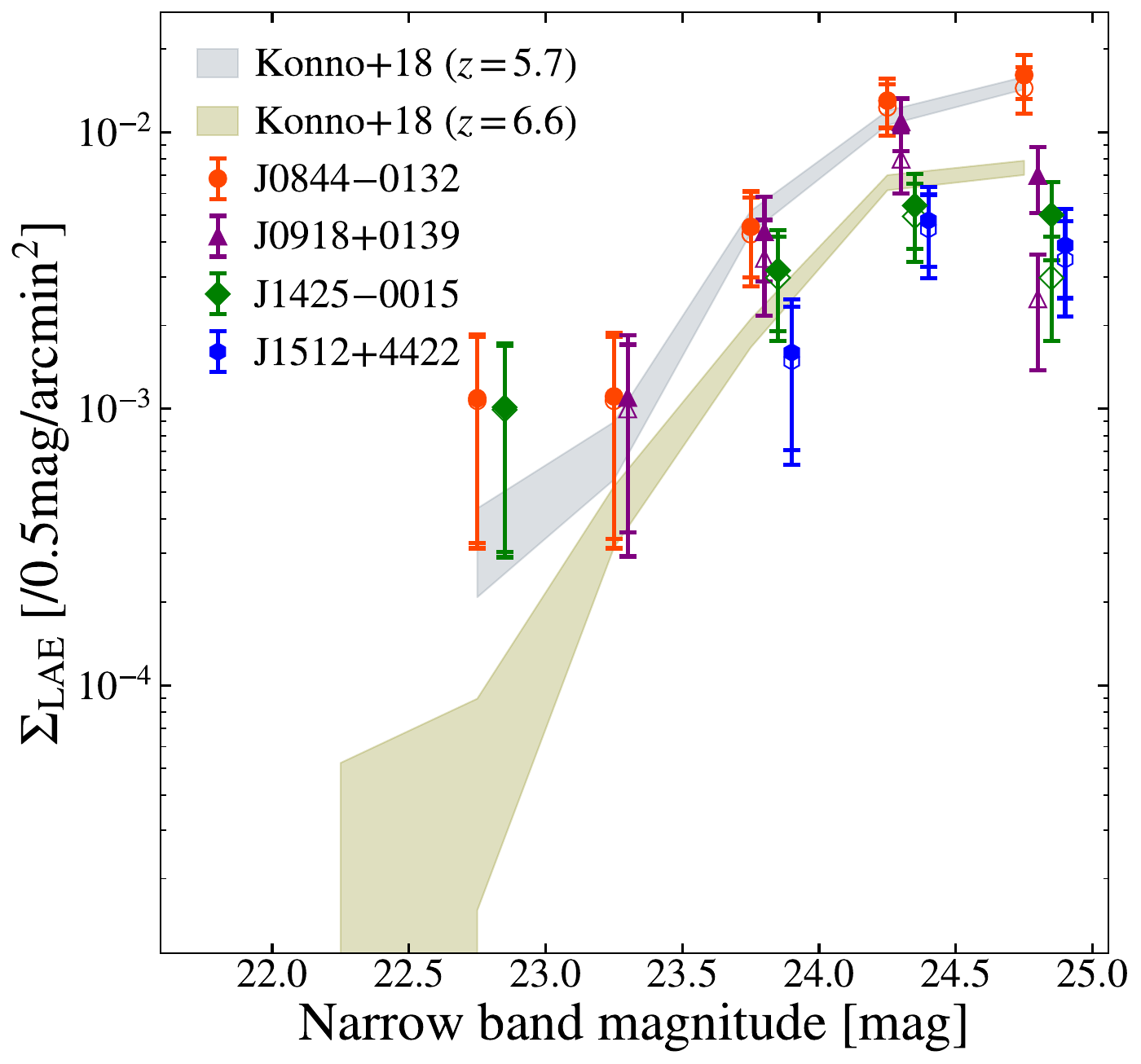}
    \caption{Surface number density of the LAEs with $r>50$ cMpc in the individual fields, where $r$ is the separation between the LAEs and the central quasars.
    The open circles show the raw counts, and the filled circles display the completeness-corrected values.
    The grey and olive shaded regions show the surface number density of LAEs with $\EW>15$~\AA\ in blank fields at $z=5.7$ and $z=6.6$ within $1\sigma$ error, respectively, as measured by \citet{Konno2018}.
    The data points are slightly offset for visibility.
    }
    \label{fig:surface_number_density_outer}
\end{figure}

We calculate the surface number density of LAEs in the individual fields.
Although all of the quasars are selected as LAEs, they are removed in the following analysis.
We correct the raw surface number densities by multiplying them by the inverse of the detection completeness in each patch.
Fig. \ref{fig:surface_number_density_outer} shows the surface number density of LAEs in the outskirts of each field, measured at projected distances greater than 50 cMpc ($\sim20\arcmin$ at $z=6.18$) from the central quasars.
This threshold, also used in a similar study by \citet{Wang2023}, is chosen to measure the LAE number density in blank fields around the large proto-clusters known at $z>6$ \citep{Jiang2018}.
Although the threshold will be much larger than the proximity zone size of our quasars, we set the value to evaluate the LAE number density in blank fields.
We compare the surface number density of LAEs in the quasar fields with that in blank fields \citep{Konno2018}, which use narrowbands of NB816 ($\lambda_c=8170$~\AA, $\mathrm{FWHM}=131$~\AA) and NB921 ($\lambda_c=9210$~\AA, $\mathrm{FWHM}=120$~\AA) on Subaru/HSC to select the LAEs with $\EW>10$~\AA\ at $z=5.7$ and $z=6.6$ in total areas of 14 and 21 deg$^2$ in the HSC-SSP Deep/UltraDeep layer, respectively.
For a fair comparison, we select LAEs with $\EW>15$~\AA\ from the LAEs in \citet{Konno2018} by applying more stringent colour criteria based on the same model in Sec. \ref{subsec:lae_selection}: $i-NB816>1.3$ for $z=5.7$ LAEs; $z-NB921>1.2$ for $z=6.6$ LAEs.
We find that the surface number densities of the fields where the effect of the quasar radiation can fully be ignored are almost comparable to those in the blank fields at similar redshifts, but the difference among the quasar fields might be caused by cosmic variance.

\subsection{Overdensity of LAEs} \label{subsec:delta}
We evaluate the overdensity of LAEs using the kernel density estimation (KDE) to calculate the LAE overdensity.
We adopt a 2D Gaussian kernel $K(\boldsymbol{X}_i, \boldsymbol{X})$;
\begin{align}
    K(\boldsymbol{X}_i, \boldsymbol{X}) = \frac{1}{2\pi b^2}\exp\left[-\frac{r(\boldsymbol{X}_i,\boldsymbol{X})^2}{2b^2}\right],
    \label{eq:kernel}
\end{align}
where $b$ is the bandwidth, and $r(\boldsymbol{X}_i,\boldsymbol{X})$ is the haversine distance between positions on the sky $\boldsymbol{X}_i$ and $\boldsymbol{X}$.
Following the approach taken in several studies (e.g., \citealp{Ito2021}) that apply the typical correlation length to the bandwidth, we set the bandwidth to the cross-correlation length of 17.7 $h^{-1}\,\mathrm{cMpc}$ ($=25.3$ cMpc) between galaxies and quasars \citep{Arita2023}, yielding $b=10\farcm5$ at $z=6.18$. 
We confirmed that the results do not change significantly even when this value is reduced to 10 cMpc.
The galaxy overdensity $\delta_\mathrm{LAE}$ at the position $\boldsymbol{X}_i$ is defined as
\begin{align}
    \delta_\mathrm{LAE}(\boldsymbol{X}_i) =\frac{\Sigma_\mathrm{LAE}(\boldsymbol{X}_i)}{\langle\Sigma_\mathrm{LAE}\rangle} - 1,
    \label{eq:delta}
\end{align}
where $\langle\Sigma_\mathrm{LAE}\rangle$ is the mean number density of LAEs in the individual fields, based on LAEs located $>50$ cMpc away from the central quasars.
The $\langle\Sigma_\mathrm{LAE}\rangle$ of the individual fields is computed by integrating the completeness-corrected surface number of LAEs for each magnitude in Fig. \ref{fig:surface_number_density_outer}.
The $\Sigma_\mathrm{LAE}(\boldsymbol{X}_i)$ is the surface number density of LAEs at the position $\boldsymbol{X}_i$, expressed as
\begin{equation}
    \Sigma_\mathrm{LAE}(\boldsymbol{X}_i)=\frac{1}{C_\mathrm{edge}(\boldsymbol{X}_i)}\cdot
    \frac{\sum_{k} w_k K(\boldsymbol{X}_k,\boldsymbol{X}_i)}{\sum_k w_k}    
\end{equation}
where $C_\mathrm{edge}(\boldsymbol{X}_i)$ is the factor to correct the edge effect due to the limited survey area and the several masked regions, and $w_k$ is the weight of the $k$th galaxy.
The sum is over the galaxies in each field.
The edge correction factor is from \citet{Jones1993}:
\begin{equation}
    C_\mathrm{edge}(\boldsymbol{X}_i) = \int_S K(\boldsymbol{X}, \boldsymbol{X}_i)dS,
    \label{eq:edge_correction}
\end{equation}
where the integration is computed over the entire unmasked survey area ($S$).
The weights are given by the inverse of the detection completeness as a function of the narrowband magnitude from Fig. \ref{fig:detection_completeness}.
Since the sensitivity drops off close to the edge of the HSC FoV.
The error of $\delta_\mathrm{LAE}$ is evaluated by propagating the Poisson statistics of the surface number densities.

\begin{figure*}
    \centering
    \includegraphics[width=2\columnwidth]{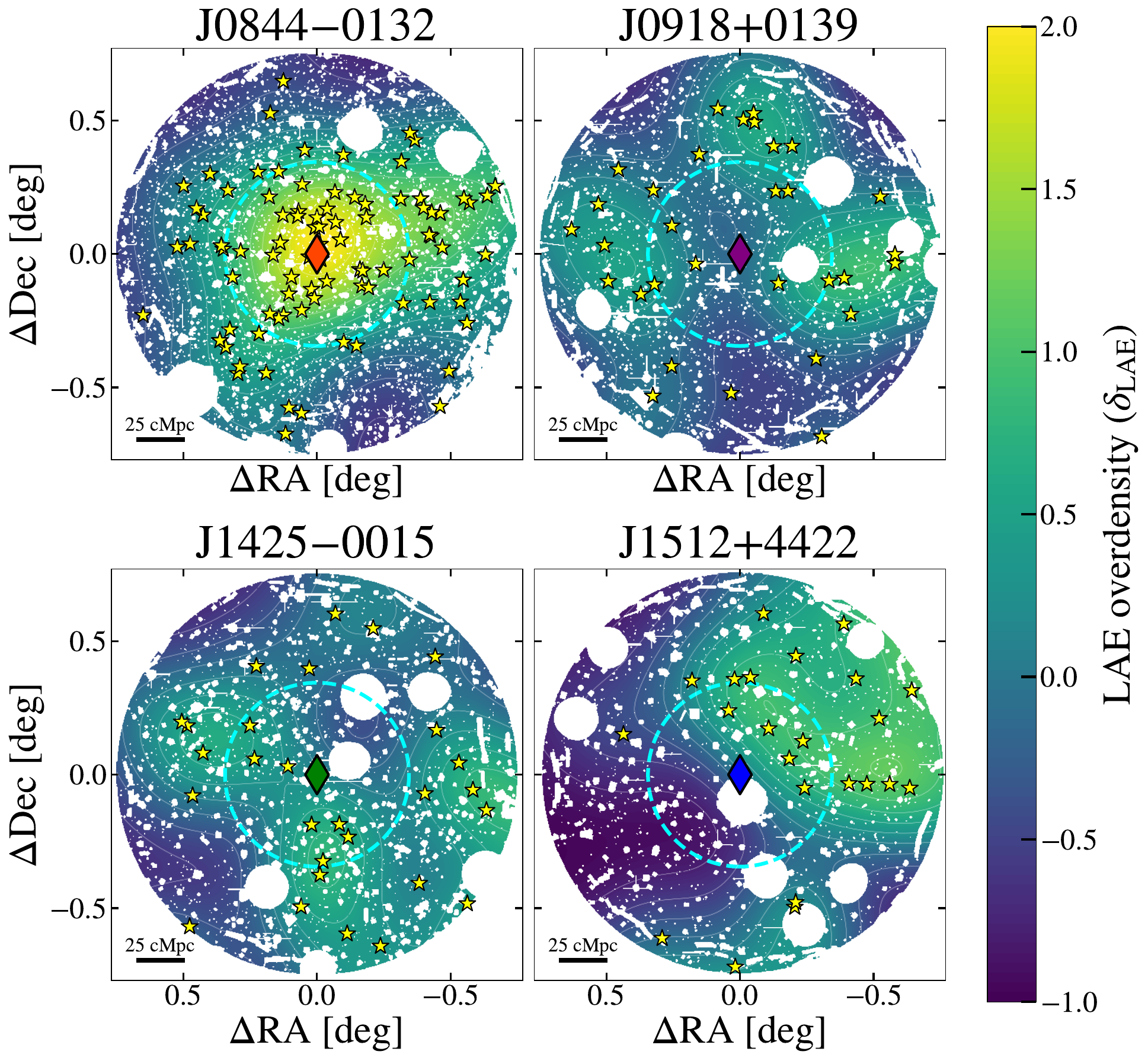}
    \caption{LAE overdensity maps of the individual fields.
    The abscissa and ordinate represent the offset from the quasar coordinate.
    The yellow stars and coloured diamonds represent the LAEs and the central quasars, respectively.
    The underlying colour and contours show $\delta_\mathrm{LAE}$ at each position calculated with equation (\ref{eq:delta}). 
    The cyan dashed lines indicate circles with a radius of 50 cMpc, beyond which the region is considered to be a blank field
    It is easier to see in Fig. \ref{fig:ew_distribution} that no LAEs are within the small proximity zones.
    }
    \label{fig:overdensity_map_corrected}
\end{figure*}

To take the sensitivity dependence on spatial position into account in the overdensity maps, we construct the completeness maps computed in Sec. \ref{subsec:lae_selection}.
We fit the detection completeness of each patch with the following function of the narrowband magnitude \citep{Serjeant2000}:
\begin{equation}
    f(m_\mathrm{AB})=\frac{f_\mathrm{max} - f_\mathrm{min}}{2}(\tanh[\alpha(m_\mathrm{AB}^{50}-m_\mathrm{AB})] + 1) + f_\mathrm{min},
    \label{eq:fcomp}
\end{equation}
where $f_\mathrm{max}$ and $f_\mathrm{min}$ are the detection completeness at the brightest and faintest magnitude, $\alpha$ quantifies the sharpness of the function, and $m_\mathrm{AB}^{50}$ is the magnitude where the detection completeness becomes 50\%.
These parameters are determined based on the $\chi^2$-fit to the detection completeness of each patch.
We calculate the detection completeness from the narrowband magnitudes of LAEs and use the reciprocal of the detection completeness as the weight to account for the spatial variation in the detection completeness.
Namely, the weight is expressed as 
\begin{equation}
    w_k=\frac{1}{f_k(NB872_k)},
\end{equation}
where $f_k$ is the fitted function of the patch containing the $k$th galaxy, and $NB872_k$ is the narrowband magnitude of the galaxy.
After this correction, the average LAE surface number densities $\langle\Sigma_\mathrm{LAE}\rangle$ are $0.015\pm0.002, 0.010\pm0.002, 0.006\pm0.001$, and $0.005\pm0.001$ arcmin$^{-2}$ for the four fields.
The difference of $\langle\Sigma_\mathrm{LAE}\rangle$ among our quasar fields is probably due to cosmic variance.
Using the average values for each of these, we calculate $\delta_\mathrm{LAE}$ of the individual fields.

Fig. \ref{fig:overdensity_map_corrected} shows the resulting LAE overdensity maps.
The inferred LAE overdensities at the quasar positions are given by $\delta_\mathrm{LAE}=1.97\pm0.40$, $-0.11\pm0.14$, $0.16\pm0.23$, and $0.01\pm0.23$ in the J0844$-$0132, J0918$+$0139, J1425$-$0015, and J1512$+$4422 fields, respectively.
They are summarized in Table \ref{tab:host_property}.
These results show that J0844$-$0132 resides in an LAE overdense region, while the other quasars do not show any significant overdensity or underdensity.
%One may expect that 
J1512$+$4422 may seem to reside at the edge of an LAE overdense region, as quasars do not always need to reside in the centre of a galaxy overdensity \citep{Zana2022, Champagne2025}.
However, it is difficult to verify the possibility based solely on the sky positions of LAEs.
Therefore, in this study, we determine whether the quasars reside in LAE overdense regions based on $\delta_\mathrm{LAE}$ at the quasar position.

\subsection{$\EW$ distribution}  \label{subsec:ew}
We estimate $\EW$ of our LAEs using the spectrum model described in Sec. \ref{subsec:lae_selection}.
Assuming that the LAEs have $\alpha=-2.5$, and their Ly$\alpha$ emission lines have a FWHM of $250\,\mathrm{km\,s^{-1}}$ at the centre of NB872, we derive the relation between $\EW$ and $z-NB872$ colour.
In our spectrum model, $z-NB872$ colour converges to $\sim2.0$ as $\EW$ approaches 240~\AA, the upper limit for the normal star-forming galaxies (SFGs) with $Z>0.02\mathrm{Z_\odot}$ \citep{Schaerer2003}.
Since our spectrum model does not consider more complex physical situations (e.g. very young galaxies, top-heavy initial mass function), which would make Ly$\alpha$ emission stronger than found in normal galaxies, and faint LAEs have a large uncertainty in $z$-band magnitudes, we assign a lower limit of $\EW=240$~\AA\ for LAEs with $z-NB872\gtrsim2.0$.
We evaluate the $\EW$ error based on the $1\sigma$ uncertainty of the $z-NB872$ colour.

\begin{figure*}
    \centering
    \includegraphics[width=2\columnwidth, clip]{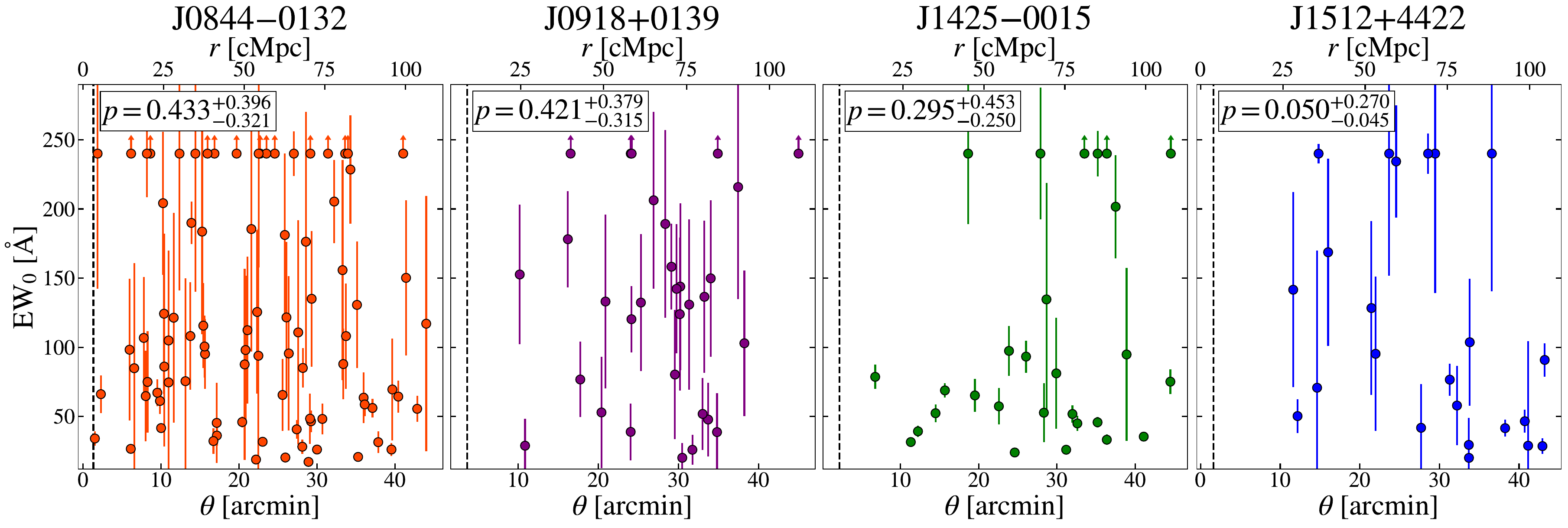}
    \caption{Relation between $\EW$ and the separation between the LAEs and the central quasars ($\theta$) of the individual quasar fields.
    The vertical dashed lines show the proximity zone sizes described in Sec. \ref{subsubsec:proximity}.
    We note that the minimum separation between J0844$-$0132 and LAEs in the quasar field is $1\farcm505$, which is sufficiently smaller than the proximity zone size of the quasar, even taking the spatial resolution of HSC into account.
    Hence, no LAEs reside within the proximity zone of the quasars.
    The left top of each panel shows the $p$-value.
    }
    \label{fig:ew_distribution}
\end{figure*}

Fig. \ref{fig:ew_distribution} shows the relation between $\EW$ of the surrounding LAEs and their angular separation ($\theta$) from the central quasars.
We verify the correlation significance by using Kendall's $\tau$ and the $p$-value under the null hypothesis that $\EW$ is not correlated with $\theta$.
We use \texttt{pymccorrelation}\footnote{\url{https://github.com/privong/pymccorrelation}} \citep{Curran2014, Privon2020}, which computes Kendall's $\tau$ and the $p$-value with censored data following \citet{Isobe1986}.
This hypothesis testing results in $p=0.433_{-0.321}^{+0.396}, 0.421_{-0.315}^{+0.379}, 0.295_{-0.250}^{+0.453}$ and $0.050_{-0.045}^{+0.270}$ for the J0844$-$0132, J0918$+$0139, J1425$-$0015, and J1512$+$4422 fields, respectively.
The errors of the $p$-values are based on the 1000 times bootstrap resampling.
The first three $p$-values suggest that the null hypothesis cannot be rejected.
While the $p$-value for the J1512$+$4422 field almost meets $p<0.05$, its large uncertainty implies that the relation between $\EW$ and $\theta$ remains merely suggestive.  
Therefore, we conclude that $\EW$ has no clear correlation with $\theta$ in any of our quasar fields.

\section{Discussion} \label{sec:discussion}

\subsection{Photoevaporation effect} \label{subsec:photoevaporation}
\subsubsection{Proximity zone size measurement} \label{subsubsec:proximity}
\begin{figure}
    \centering
    \includegraphics[width=\columnwidth]{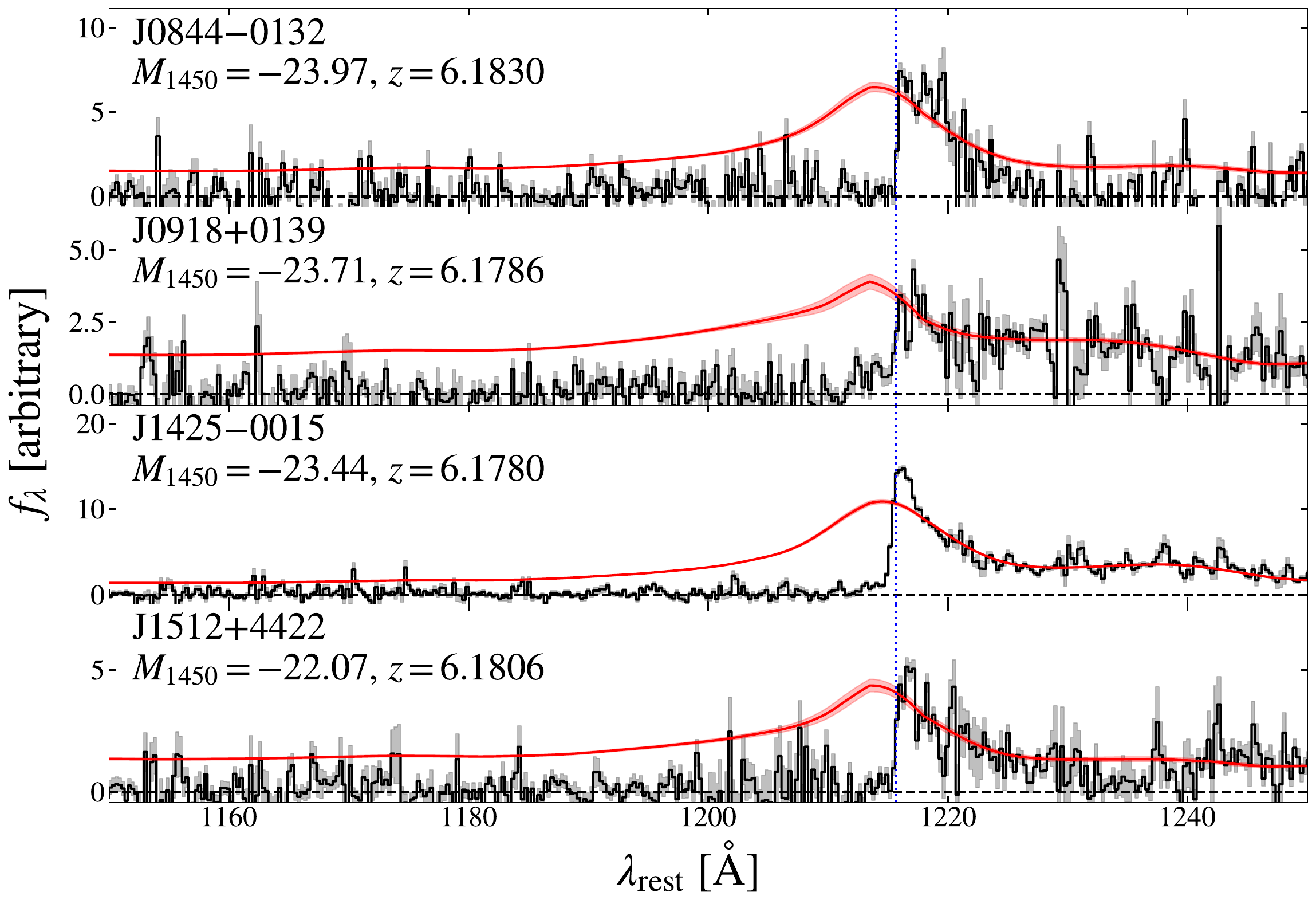}
    \caption{
    Observed quasar spectra and the quasar continua predicted by QFA.
    The black lines with the grey shade show the observed quasar spectra with their errors normalized by the flux at 1280~\AA\ in the rest frame.
    The red lines and shade represent the QFA-predicted quasar continua and their uncertainty.
    The blue dotted lines indicate the quasar redshift.
    }
    \label{fig:qfa_continuum}
    \includegraphics[width=\columnwidth, clip]{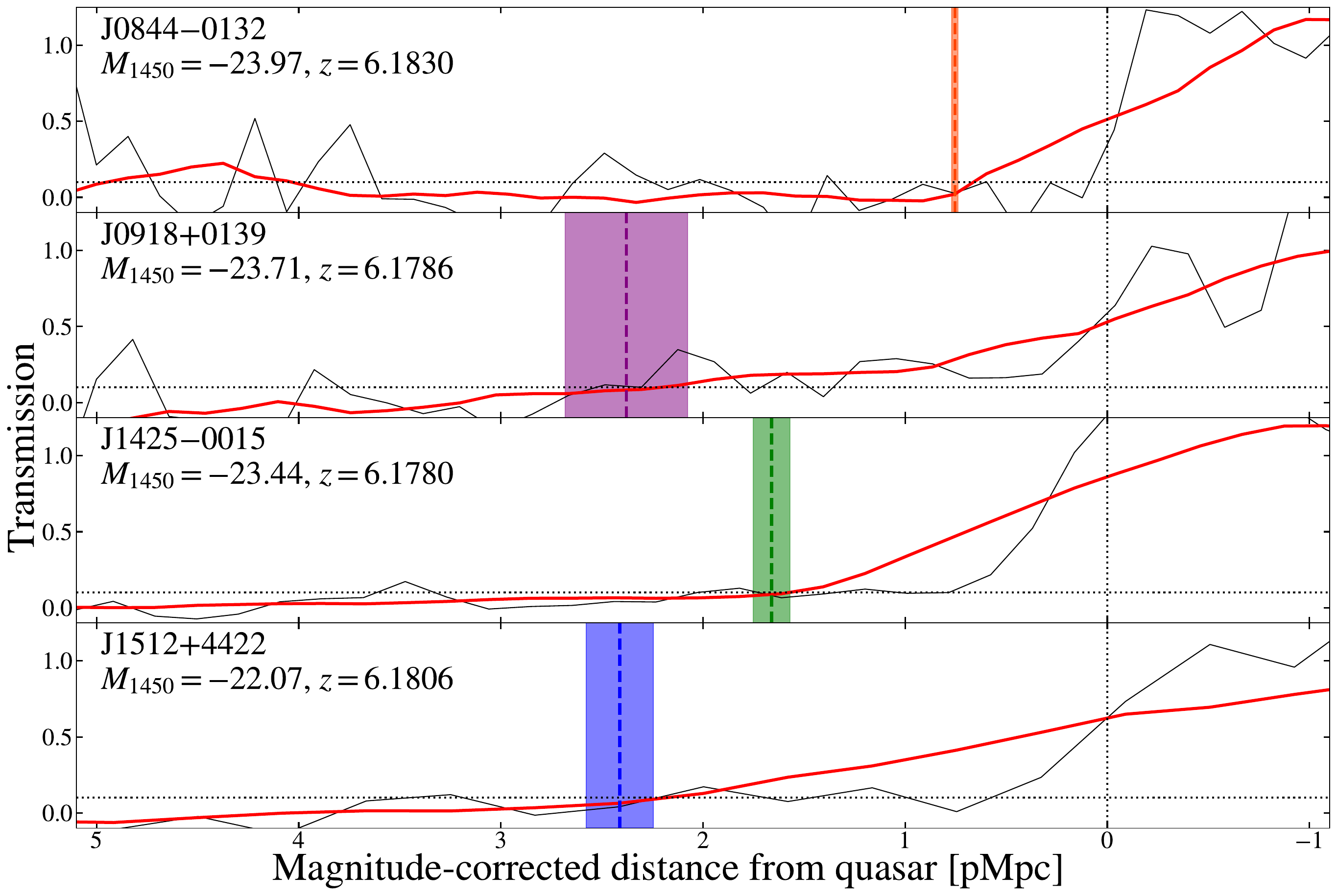}
    \caption{
    IGM transmission as a function of the magnitude-corrected distance using equation(\ref{eq:magnitude_correction}) from our quasars.
    The black lines show the IGM transmission based on the continuum recovered by QFA.
    The red lines represent the IGM transmission smoothed with a 20~\AA\ wide boxcar function.
    The horizontal dotted lines show a transmission level of 10\%.
    The vertical dotted lines denote the systemic redshift of the quasars.
    The dashed lines with the shaded regions show $R_\mathrm{p,\,corr}^{-25}$ and its error.
    }
    \label{fig:qfa_transmission}
\end{figure}

We measure the proximity zone sizes of our quasars using their FOCAS spectra
\footnote{The spectra of our quasars are published on \url{https://cosmos.phys.sci.ehime-u.ac.jp/~yk.matsuoka/shellqs.html}.}
.
The spectral coverage is from 0.75 \micron\ to 1.05 \micron\ with a resolution $R\sim1200$.
First, we infer the intrinsic quasar spectra using an unsupervised probabilistic model, named Quasar Factor Analysis (QFA\footnote{\url{https://github.com/ZechangSun/QFA}}; \citealp{Sun2023}).
QFA predicts the intrinsic quasar continua based on the latent factor analysis, which is a statistical framework that assumes that a high-dimensional correlated data set (the quasar continua) can be expressed as linear combinations of a small set of lower-dimensional latent factors (e.g. Ly$\alpha$ emission, power-law feature of quasar continua, \ion{C}{IV} emission), combined with the physical priors of the IGM.
\citet{Sun2023} demonstrate that QFA achieves lower absolute fractional flux error in the recovered quasar continua than does a conventional principal component analysis-based method.
QFA uses the quasar spectra normalized by the flux at 1280~\AA\ in the rest frame, the flux uncertainty, and the systemic redshift determined by [\ion{O}{III}]$\lambda5007$ emission in Table \ref{tab:qso_properties}.
When applying QFA to our quasars' spectra, we conform the observed spectra onto a uniform rest-frame wavelength grid from 1030~\AA\ to 1600~\AA\ with a logarithmic uniform spacing (median pixel size is $\sim0.3$~\AA, corresponding to $\sim0.014$ pMpc) and mask wavelength ranges where the normalized flux shows low signal-to-noise ratios.
The QFA-predicted quasar continua are shown in Fig. \ref{fig:qfa_continuum} with the observed quasar spectra.

We measure the proximity zone size following the standard definition used in the literature (e.g. \citealp{Fan2006, Carilli2010, Venemans2015, Eilers2017, Ishimoto2020}).
We smooth the quasar spectra with a 20~\AA\ wide (observed frame) boxcar function.
The smoothing scale corresponds to 0.93 pMpc at $z=6.18$.
The IGM transmission is computed as the ratio of the smoothed quasar spectra and the smoothed QFA-predicted continua.
The proximity zone size is defined as the distance to the first of three sequential pixels that show transmission below 10\% blueward of the Ly$\alpha$ emission line.
The error of the proximity zone size is estimated based on the observed error in the quasar spectra.
After randomly adding pixel-to-pixel errors to the quasar spectra, we repeat the measurements of the proximity zone size 1000 times, and consider the variation in these measurements as the error of the proximity zone size.
We obtain $R_\mathrm{p}=0.4449\pm0.0093,1.230\pm0.156,0.7478\pm0.0404,0.5387\pm0.0373$ pMpc for J0844$-$0132, J0918$+$0139, J1425$-$0015, J1512$+$4422, respectively.
For a fair comparison later, we also compute the luminosity-corrected proximity zone size based on $M_{1450}$ following \citet{Ishimoto2020}, which is applicable to low-luminosity regimes such as our sample:
\begin{equation}
    R_\mathrm{p,\,corr}^{-25} = R_\mathrm{p}\times 10^{0.4\times(M_{1450}+25)/1.80}.
    \label{eq:magnitude_correction}
\end{equation}
The corrected proximity zone sizes are $R_\mathrm{p,\,corr}^{-25}=0.7536\pm0.0157,2.379\pm0.302,1.661\pm0.090,2.412\pm0.167$ pMpc for J0844$-$0132, J0918$+$0139, J1425$-$0015, J1512$+$4422, respectively.
These results are summarized in Table \ref{tab:host_property}.
Fig. \ref{fig:qfa_transmission} shows the IGM transmission based on the quasar continua recovered by QFA.

\begin{table*}
    \centering
    \caption{Physical properties of host galaxies and the SMBHs.
    The $M_*$ are from \citet{Ding2023, Ding2025}, the SFRs are from \citet{Phillips2025}, and the SMBH masses and the bolometric luminosities are from Onoue et al. in prep. We note that the SMBH masses are based on the H$\beta$ emission lines.
    }
    \label{tab:host_property}
    \begin{tabular*}{2\columnwidth}{@{\extracolsep{\fill}}cccccc} \hline
        Field & $\langle\Sigma_\mathrm{LAE}\rangle$ & $\delta_\mathrm{LAE}$ & $R_\mathrm{p}$ & $R_\mathrm{p}$ & $R_\mathrm{p,\,corr}^{-25}$\\
         & (arcmin$^{-2}$) & & (arcmin) & (pMpc) & (pMpc) \\ \hline
        J0844$-$0132 & $0.015\pm0.002$ & $1.97\pm0.40$ & $1.320\pm0.027$ & $0.4449\pm0.0093$ & $0.7536\pm0.0157$\\
        J0918$+$0139 & $0.010\pm0.002$ & $-0.11\pm0.14$ & $3.646\pm0.462$ & $1.230\pm0.156$ & $2.379\pm0.302$ \\
        J1425$-$0015 & $0.006\pm0.001 $ & $0.16\pm0.23$ & $2.218\pm0.120$ & $0.7478\pm0.0404$ & $1.661\pm0.090$\\
        J1512$+$4422 & $0.005\pm0.001$ & $0.01\pm0.23$ & $1.598\pm0.111$ & $0.5387\pm0.0373$ & $2.412\pm0.167$\\ \hline
    \end{tabular*} \\
    \vspace{2mm}
    \begin{threeparttable}
    \begin{tabular*}{2\columnwidth}{@{\extracolsep{\fill}}ccccc} \hline
        Stellar mass & SFR & sSFR & SMBH mass & Bolometric luminosity \\
         $\log(M_*/\Msun)$ & ($\Msun\,\mathrm{yr^{-1}}$) & (Gyr$^{-1}$) & $\log(M_\mathrm{BH}/\Msun)$ & $\log(L_\mathrm{bol}/\mathrm{erg\,s^{-1}})$ \\ \hline
        $10.03_{-0.45}^{+0.53}$ & $39_{-24}^{+9}$ & $4_{-5}^{+9}$ & $8.57$ $^{\dagger}$ & $46.427\pm0.001$ $^{\dagger}$ \\
        $10.21_{-0.37}^{+0.52}$ & -- & -- & $8.57$ & $45.827\pm0.003$ \\
        $10.54_{-0.37}^{+0.36}$ & $37_{-27}^{+5}$ & $0.8_{-0.6}^{+1.2}$ & $8.43$ & $45.976\pm0.003$ \\
        $10.57_{-0.41}^{+0.29}$ & $33_{-16}^{+4}$ & $0.7_{-0.4}^{+1.2}$ & $9.17$ & $45.728\pm0.002$ \\ \hline
    \end{tabular*} 
    \begin{tablenotes}
        \item[$\dagger$] Onoue et al. in prep. present the SMBH mass and the bolometric luminosity without correction for dust extinction, although \citet{Matsuoka2025} show that J0844$-$0132 is a dust-obscured quasar with $A_V\sim2$ based on the Balmer decrement.
        \citet{Matsuoka2025} present the SMBH mass and bolometric luminosity corrected for dust extinction as $\log(M_\mathrm{BH}/\Msun)=9.11\pm0.01$ and $\log(L_\mathrm{bol}/\mathrm{erg\,s^{-1}})=47.08\pm0.01$.
    \end{tablenotes}
    \end{threeparttable}
\end{table*}

\subsubsection{Does photoevaporation affect LAE overdensity?}
\label{subsubsec:photoevaporation}
We find that no LAEs reside in the proximity zones in any of our quasar fields, as shown in Fig. \ref{fig:ew_distribution}.
Despite being an LAE overdense region, the J0844$-$0132 field contains no LAEs within its proximity zone.
Applying the $\Sigma_\mathrm{LAE}$ at the quasar position measured in Section \ref{subsec:delta} yields an expected LAE detection of 0.12 within the proximity zone size. 
Hence, we cannot conclude that the photoevaporation effect hinders the formation of LAEs.
On the other hand, although the proximity zone size of our quasars is quite small compared to the expected overdensity scale, LAE overdensities are not observed in the remaining three fields either.
This allows us to conclude that, at least in these three regions, the absence of LAE overdensity is not due to photoevaporation effects from the quasars.

In estimating the proximity zone size above, it is assumed that quasar radiation is isotropic. 
We caution that the result may not hold if the quasar radiation is beamed in a specific direction, but only the sky positions of LAEs are not sufficient to verify this possibility because the effect is diluted due to the projection effect; therefore, it is necessary to unveil the three-dimensional distribution of the LAEs by spectrocopic confirmation.
The proximity zone size also depends on the quasars' age, which will be discussed later.
\citet{Uchiyama2019} and \citet{Suzuki2025} suggest that LAEs with high Ly$\alpha$ $\EW$ are particularly scarce around quasars, as these LAEs are typically low-mass galaxies vulnerable to the photoevaporation effect. 
However, in our four quasar fields, we found no statistically significant correlation between $\EW$ and $\theta$ (Fig. \ref{fig:ew_distribution}).
Our results, which show a wide range of environments from $\delta_\mathrm{LAE}=-0.11$ to $\delta_\mathrm{LAE}=1.97$, indicate that the LAE density around the quasar exhibits diversity independent of the photoevaporation effect.

The absence of LAEs within the proximity zone is also observed in the bright quasar VIK J2348$-$3054 at $z=6.90$ \citep{Lambert2024}, located within an LAE overdensity over several pMpc.
\citet{Lambert2024} suggest that the absence is caused by the photoevaporation effect.
On the other hand, several LAEs are found in the proximity zones of some quasars at $z\sim6$ (e.g. \citealp{Bosman2020, Protusova2025}).
The presence of these LAEs is thought to be due to the effect of locally small \ion{H}{I} fractions caused by ionizing photons from quasars rather than photoevaporation.

Quasars with a very small proximity zone size ($R_\mathrm{p,\,corr}^{-25}\lesssim0.90$ pMpc) are suggested to be very young ($\lesssim10^4$ yr) \citep{Eilers2017, Ishimoto2020}.
Since it takes time for the IGM to reach ionization equilibrium, the small proximity zone size indicates that the duration of the quasar activity has been relatively small, as confirmed by a radiative transfer simulation by \citet{Davies2020}.
Interestingly, the only quasar that meets this crierion for being a young quasar is J0844$-$0132, which resides in an overdense region, meets the criterion of a young ($\lesssim10^4$ yr) quasar among our quasars.
The proximity zone sizes of the remaining three quasars are consistent with the relation between $R_\mathrm{p}$ and $M_{1450}$ derived in \citet{Ishimoto2020}, which implies that these quasars are not extraordinarily young.
These results suggest that the quasar age could be related to the LAE overdensity in the quasar fields.
One interpretation of this relation could be that as long as the quasar is young, its intense radiation would not have had sufficient time to suppress even nearby galaxies $<1$ pMpc away from the quasar, even if the quasar is sufficiently luminous to hinder their star formation by the photoevaporation effect.
However, the current data are too limited to draw strong conclusions, and the scenario needs to be verified with a larger sample.
Moreover, J0844$-$0132 is a mildly obscured quasar \citep{Matsuoka2025}, and in this case, the number of ionizing photons escaping from the quasar decreases, resulting in a smaller apparent proximity zone size. 
Consequently, the quasar age estimated from this measurement may be underestimated.
It also should be noted that foreground \ion{H}{I} clouds, such as damped Ly$\alpha$ systems and Lyman Limit systems, may affect the proximity zone size measurements.
To determine if this is an effect, deeper spectroscopic follow-up observations are necessary.

\subsection{Relation between LAE overdensity and the properties of the host galaxy and SMBH} \label{subsec:quasar_properties}
We discuss whether the physical properties of the quasars are related to the LAE overdensity in their fields.
In order to investigate the relation between the various quasar properties and the LAE overdensity, we include seven luminous quasars at $5.7<z<7$ from previous studies in Table \ref{tab:previous_studies} in addition to our low-luminosity quasars.
The effective area of the previous studies listed in Table \ref{tab:previous_studies} varies from VLT/MUSE ($\sim1$ arcmin$^2$) to DECam ($\sim3$ deg$^2$), depending on the FoV of the instrument used.
Four of the luminous quasars (J1030$+$0524, J0836$+$0054, J0910$-$0414, J2348$-$3054) are in LAE overdense regions, while three (J0203$+$0012, J2329$-$0301, J0305$-$3150) are in non-overdense regions.

\subsubsection{Host galaxy properties} \label{subsubsec:host_galaxy}
\begin{figure}
    \centering
    \includegraphics[width=\columnwidth,clip]{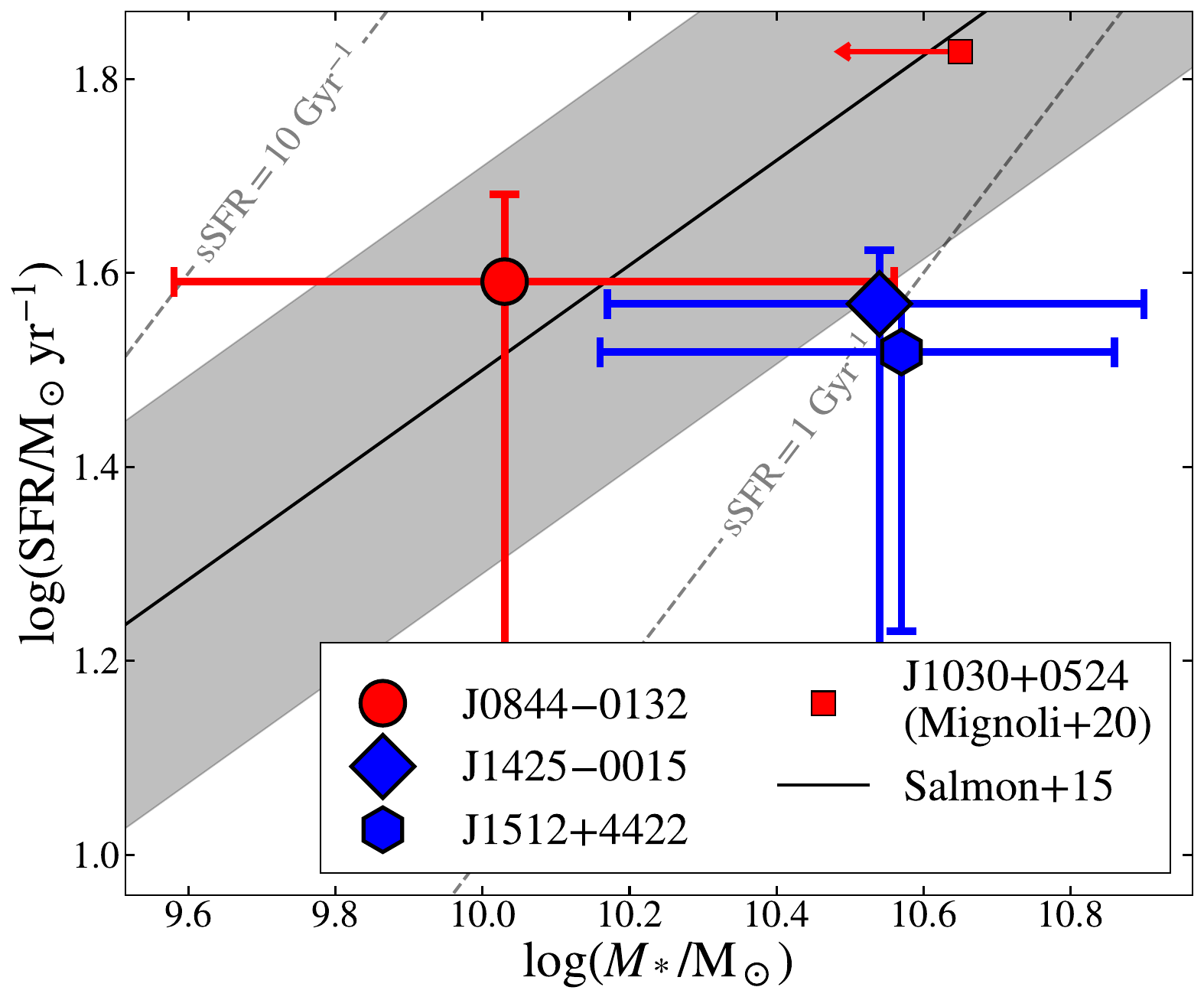}
    \caption{
    The relationship between $M_*$ and the SFR of the quasars in this study and previous research.
    The red and blue symbols show that the quasars reside in the LAE overdense and non-overdense regions, respectively.
    The dashed lines represent $\mathrm{sSFR}=10\,\mathrm{Gyr^{-1}}$ (top) and $\mathrm{sSFR}=1\,\mathrm{Gyr^{-1}}$ (bottom).
    The solid line with the grey shade shows the star-formation main sequence with its uncertainty at $z\sim6$ \citep{Salmon2015}.
    The $M_*$ of our quasars are from \citet{Ding2023, Ding2025}, and the SFRs and sSFRs are from \citet{Phillips2025}.
     The SFR and sSFR of J0918$–$0139 have not been measured, as no spatially extended H$\alpha$ emission has been detected.
    The $M_*$ and SFR of J1030$+$0524 are from \citet{Yue2024} and \citet{Decarli2018}, and \citet{Mignoli2020} confirm that the quasar resides in an LAE overdense region.
    Due to the extreme brightness of J1030$+$0524, they do not detect its host galaxy in all the PSF-subtracted images, so they only constrain the upper limit of the $M_*$.
    The SFR is derived from the FIR luminosity following \citet{Kennicutt2012}.
    }
    \label{fig:LAE_overdensity_host}
\end{figure}

Since $M_*$ and DMH mass are positively correlated \citep{Behroozi2013, Behroozi2019}, one might expect that quasars hosted by galaxies with high $M_*$ will be found in galaxy overdensities.
The host galaxies of our quasars have been directly detected by \textit{JWST}, and their $M_*$ and SFR have been measured, which allows us to test this scenario.
We examine the correlation between the LAE overdensity and these host galaxy properties, which are summarized in Table \ref{tab:host_property}.
The SFR of J0918$+$0139 is not constrained because its extended H$\alpha$ emission line is not confirmed.
The 10 Myr-scale SFR of J1512$+$4422 is estimated by \citet{Onoue2025} and \citet{Phillips2025}, but the latter estimates a $\sim1$ dex higher value than the former.
To infer its SFR, the former performs the spectral energy distribution (SED) fitting to its quasar-decomposed spectrum; so SFR averaged over 100 Myr, while the latter uses the extended H$\alpha$ emission (SFR over 10 Myr) based on the assumption that it is due to the star formation activity in the host galaxy and that the specific SFR .
For consistency of the SFR estimate in our quasars, we adopt the SFR by \citet{Phillips2025} for all of our quasars in the following discussion.
We confirm that adopting the SFR for J1512$+$4422 in \citet{Onoue2025} does not have an impact on the results that the SFRs are not strongly correlated to the LAE overdensity.
Another luminous quasar, J1030$+$0524, known to reside within a LAE high-density region \citep{Mignoli2020}, has a upper limit of $\log(M_*/\Msun)<10.65$ constrained with the NIRCam imaging with F115W, F200W, and F356W \citep{Yue2024} and its SFR derived from the far-infrared (FIR) luminosity measured by ALMA observations \citep{Decarli2018} as $67\,\Msun\,\mathrm{yr^{-1}}$.
We caution that the SFR estimated from FIR luminosity averages star formation activity over 100 Myr, which is different from the 10 Myr timescale of H$\alpha$-based SFR \citep{Kennicutt2012}.
We note that this is the only quasar in Table \ref{tab:previous_studies} whose host properties have been measured.
The table summarizes the previous studies on the LAE overdensity around quasars at $6\lesssim z \lesssim7$.
The LAE overdensity measurement can vary with the limiting magnitude of the sample and the measured spatial range, making quantitative comparisons with past studies difficult. 
Therefore, we will not consider the significance of overdensity here, but only compare whether overdensity is present or not.
Fig. \ref{fig:LAE_overdensity_host} shows the relation between the LAE overdensity and the $M_*$, SFR, sSFR of the host galaxies.
We find that J0844$-$0132, which exhibits LAE overdensity in our quasars, shows the smallest $M_*$ and the largest sSFR among our quasars, though the differences are smaller than the uncertainties as shown in \citet{Phillips2025}.
All of the host galaxies with SFR available in our quasars lie close to the star-forming main sequence at $z\sim6$ \citep{Clarke2024}.
We conclude that LAE overdensity does not show clear correlations with the host galaxy properties.
For reference, Fig. \ref{fig:mstar_sfr_appendix} shows the relationship between host galaxy properties and surrounding overdensity, including results from galaxy populations other than LAEs. 
In this case, one must be very careful as the galaxies used to measure overdensity, the effective area over which they are measured, and the methods used to measure SFR and $M_*$ differ. 
However, no clear correlations are observed.

At low-$z$, $M_*$ is correlated with DMH mass \citep{Moster2010}, and thus should be correlated with surrounding galaxy overdensity.
However, our results showed no such tendency.
\citet{Ding2025} show that the host galaxies of our quasars have high $M_*$, lying at the massive end of the stellar mass function at $z\sim6$, but only J0844$-$0132 resides in an LAE overdense field.
This may be because there is a large dispersion in the stellar-to-halo mass ratio of the high-$z$ quasars in addition to the narrow dynamic range of $M_*$ of this sample.
Further observations are necessary to investigate the relation between the LAE overdensity and the host $M_*$.

\subsubsection{SMBH properties} \label{subsubsec:smbh}
\begin{figure}
    \centering
    \includegraphics[width=\columnwidth,clip]{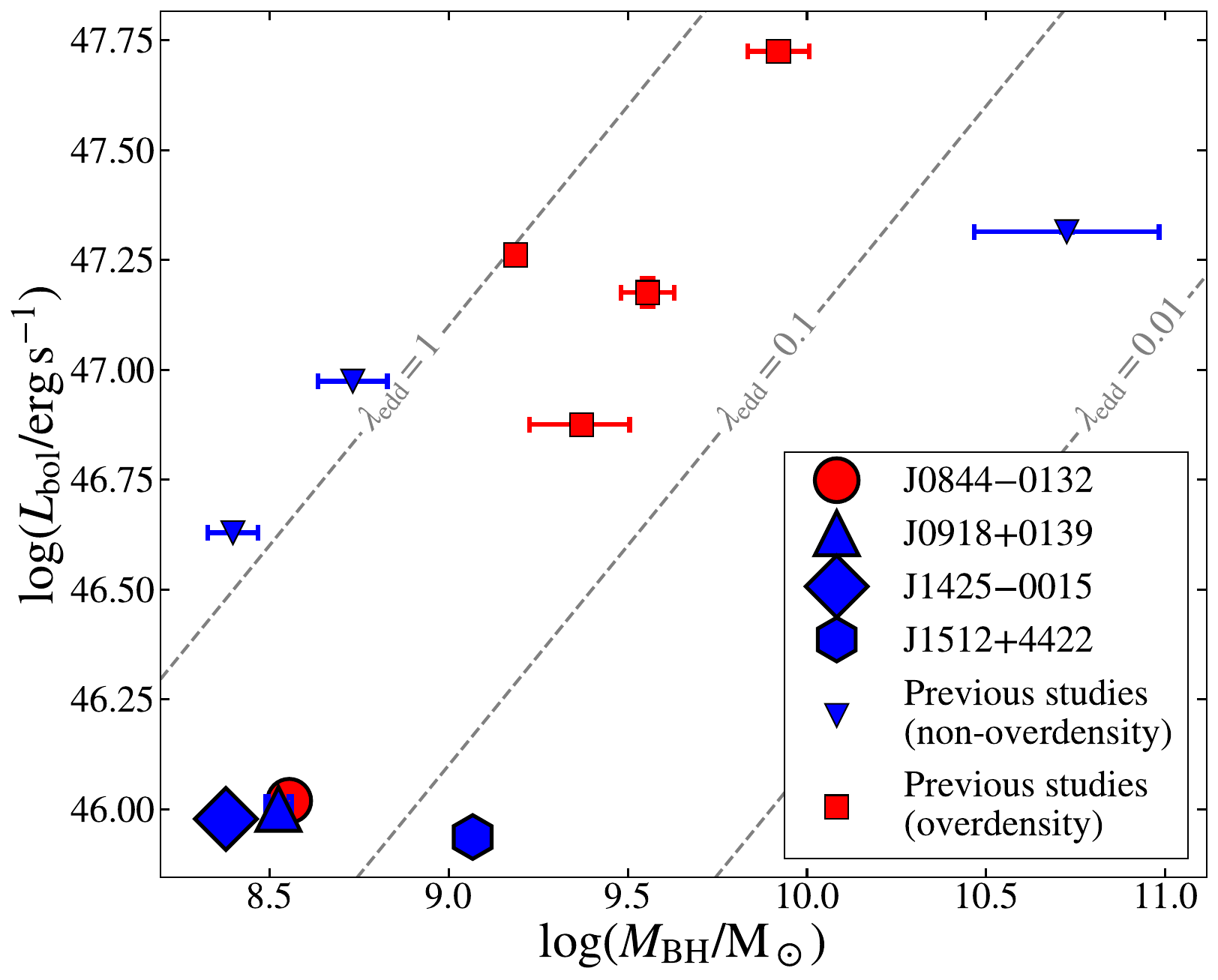}
    \caption{
    The SMBH masses and the bolometric luminosities of the quasars in this study and literature.
    The red and blue symbols show the quasars that reside in regions overdense and non-overdense in LAEs, respectively.
    The dashed lines show the Eddington ratio, $\lambda_\mathrm{Edd}=1, 0.1, 0.01$ from top to bottom.
    These SMBH properties of our quasars are from Onoue et al. in prep., which estimate the SMBH mass based on the H$\beta$ emission \citep{Kaspi2000}.
    We also plot results of the quasars in Table \ref{tab:previous_studies} from the literature.
    We use those parameters in \citet{Shen2019} for J0203$+$0012 and J0836$+$0054, \citet{Willott2010} for J2329$-$0301, \citet{Farina2022} for  J0305$-$3150 and J2348$-$3054, \citet{Yang2021} for J0910$-$0414, and \citet{Yue2024} for J1030$+$0524.
    \citet{Shen2019} estimate the SMBH mass based on the \ion{C}{IV} $\lambda1549$ line \citep{Vestergaard2006} and compute the bolometric luminosity based on the continuum luminosity at 3000~\AA\ in rest-frame with a bolometric correction of 5.15 \citep{Richards2006}.    
    \citet{Willott2010} estimate the SMBH mass based on the \ion{Mg}{II} $\lambda2798$ line \citep{Vestergaard2009} and compute the bolometric luminosity based on the continuum luminosity at 3000~\AA\ in rest-frame with a bolometric correction of 6 \citep{Richards2006, Jiang2006}.
    \citet{Farina2022} and \citet{Yang2021} estimate the SMBH mass based on the \ion{Mg}{II} $\lambda2798$ line \citep{Vestergaard2009} and the bolometric luminosity from the continuum luminosity at 3000~\AA\ in rest-frame with a bolometric correction of 5.15 \citep{Richards2006}.
    \citet{Yue2024} estimate the SMBH mass based on the H$\beta$ line \citep{Vestergaard2006} and the bolometric luminosity from the continuum luminosity at 5100~\AA\ in rest-frame with a bolometric correction of 9.26 \citep{Runnoe2012}.
    We caution that we use the available SMBH mass measurements in the literature, which induce the SMBH mass measured by different methods.
    }
    \label{fig:LAE_overdensity_smbh}
\end{figure}

We also compare the LAE overdensity of our quasars with their SMBH properties (Onoue et al. in prep.).
Fig. \ref{fig:LAE_overdensity_smbh} shows the comparison of SMBH mass and the bolometric luminosity among quasars whose LAE overdensity has been measured.
We note that \citet{Matsuoka2025} present SMBH mass and bolometric luminosity of J0844$-$0132 $\sim0.6$ dex higher than those from Onoue et al. in prep., because they applied a correction for dust extinction based on the observed Balmer decrement, yielding $A_V=2.3$.
However, in this case, applying the standard SMC dust extinction law \citep{Gordon2003} would result in an extinction of $>10$ mag in the Ly$\alpha$ emission; yet the observed Ly$\alpha$ emission has highly prominent luminosity, suggesting that this extinction correction is overestimated.
Hence, we adopt these SMBH properties without the correction.
The following discussion is not affected by which values are applied.
No SMBH parameters appear to be related to the LAE overdensity in Fig. \ref{fig:LAE_overdensity_smbh}.
These results suggest that the SMBH properties are not significantly related to their large-scale environment.
This is consistent with the recent results for low-luminosity AGNs at $3.9<z<6$ \citep{Lin2025}, which investigate the large-scale environments of 28 low-luminosity AGNs using 782 H$\alpha$ emitters (HAEs) to find no clear correlation between the surrounding HAE overdensity and the AGN properties (broad-line luminosity, BH mass, AGN fraction).
For reference, Fig. \ref{fig:mbh_lbol_appendix} shows the relationship between SMBH properties and surrounding overdensity for samples from the literature, including recent results from EIGER and ASPIRE.
One must be very careful since the galaxy populations used to measure overdensity, and the effective area over which they are measured differ, but no clear correlation is observed.

\begin{table*}
    \centering
    \caption{Previous studies on the LAE overdensity over Mpc-scale around luminous quasars at $5.7< z<7$.
    The effective area in this table represents the area to search for LAEs homogeneously.
    }
    \label{tab:previous_studies}
    \begin{threeparttable}
    \begin{tabular}{lccccccp{3.8cm}}\hline
       Quasar & $z$ & $M_{1450}$ &  Effective area & Overdensity &  Redshift ref. & Overdensity ref. & Comments \\ 
        & & (mag) & (arcmin$^2$) & &  &\\ \hline
       ULAS J020332.38$+$001229.2 & $5.72$ & $-26.2$ &  $44$ & No  & (1) &  (8) &  \\ 
       PSO J215.1512$-$16.0417 & $5.73$ & $-27.6$  & $37$ & No & (2)  & (9) & No $M_\mathrm{BH}$ measurement\\
       SDSS J083643.85$+$005453.3 & $5.81$ & $-27.8$ &  $11$  & Yes & (3)  & (10) & \\
       SDSS J103027.10$+$052455.0 & $6.31$ & $-27.1$ &  $1$ & Yes$^{\ddagger}$ &  (4) & (11) & $M_*$ and SFR have been measured in \citet{Yue2024} and \citet{Decarli2018}, respectively.\\
       CFHQS J232908$-$030158.8 & $6.42$ & $-25.2$ &  $788$ & No$^\dagger$  & (5) & (12) & \\
       VIKING J030516.92$-$315056.0 & $6.61$ & $-26.0$  & $697$ & No & (6) & (13) & \\
       DELS J091054.53$-$041406.8 & $6.63$ & $-26.4$  & $3740$ & Yes & (7) & (14) & \\
       VIKING J234833.34$-$305410.0 & $6.90$ & $-25.7$  & $10332$ & Yes & (4) & (15) & \\
       \hline
    \end{tabular}
    \begin{tablenotes}
        \item References: 
        (1) \citet{Venemans2007}; 
        (2) \citet{Banados2014}; 
        (3) \citet{Banados2016};
        (4) \citet{Kurk2007}; 
        (5) \citet{Willott2007}; 
        (6) \citet{Venemans2013};
        (7) \citet{Wang2019b}; 
        (8) \citet{Banados2013}; 
        (9) \citet{Mazzucchelli2017}; 
        (10) \citet{Overzier2022}; 
        (11) \citet{Mignoli2020}; 
        (12) \citet{Goto2017}; 
        (13) \citet{Ota2018}; 
        (14) \citet{Wang2024}; 
        (15) \citet{Lambert2024}
        \item[$\dagger$] \citet{Farina2017} discovered an LAE by VLT/MUSE at $\sim12.5$ pkpc and $560\,\mathrm{km\,s^{-1}}$ from CFHQS J232908$-$030158.8, suggesting that the quasar resides in the high-density environment.
        However, the FoV of MUSE is much smaller than that of Subaru/Suprime-Cam used in \citet{Goto2017}.
        This study focuses on the LAE distribution in the high-$z$ quasar fields with a large FoV, so we include CFHQS J232908$-$030158.8 in the quasars not residing in an LAE overdense field.
        \item[$\ddagger$] \citet{Balmaverde2017} presented 21 robust LBGs in this quasar field ($\sim25\arcmin\times25\arcmin$) using multiband photometry of $Y,J$-bands from WIRCam on Canada-France-Hawaii Telescope and literature ($r,i,z$-bands: \citealp{Morselli2014}; $H, K$-bands: \citealp{Quadri2007, Blanc2008}; Spitzer Infrared Array Camera 3.4 and $4.5\micron$ bands: Spitzer Heritage Archive), and \citet{Mignoli2020} performed spectroscopic follow-up observations for 12 objects to find that nine of them are real galaxies at $z>5.7$.
        They reported that four galaxies are associated with the large-scale structure of SDSS J010013.02+280225.8 and that three of them show $\EW\gtrsim15$~\AA, which can be regarded as LAEs.
        Therefore, while we treat CFHQS J232908$-$030158.8 as a quasar in a non-overdense region, we count SDSS J103027.10$+$052455.0 among quasars in overdense regions.
    \end{tablenotes}
\end{threeparttable}
\end{table*}

On the other hand, theoretical studies suggest a correlation between the galaxy overdensity and the SMBH properties.
\citet{Pizzati2024} predicts that more luminous quasars should lie in more massive DMHs at $z\sim6$ by combining a large cosmological simulation and observed luminosity function and correlation function of the quasars
However, the scatter of their predicted relation is as large as $\sim0.64$ dex.
The correlation is a consequence of the fact that the massive DMH can accumulate a large amount of gas, which then accretes onto the SMBH.
This implies that luminous quasars are expected to lie in galaxy overdense regions.
\citet{Habouzit2019} explore the expected relation between the galaxy overdensity and the SMBH mass at high-$z$.
They employ the large-scale simulations of Horizon-AGN \citep{Dubois2014} and Horizon-noAGN \citep{Peirani2017} to show that SMBHs with $M_\mathrm{BH}\geq10^8\,\Msun$ at $z\sim6$ are embedded in more overdense regions within 10 cMpc than galaxies with the same $M_*$ without SMBHs, and that the number of surrounding galaxies increases with the SMBH mass.
One of the explanations for the inconsistency might be the projection effect.
While the three-dimensional distance between galaxies is available in the cosmological simulation, the narrowband observation only provides the projected distance among LAEs, which can dilute the overdensity significance due to the contamination by field galaxies.
In fact, the FWHM of NB872 corresponds to 22 cMpc along the line-of-sight, which is larger than the expected overdensity scale of $\sim10$ cMpc \citep{Habouzit2019}.
Another possible reason for the discrepancy is that our sample is limited to only four quasars, and that these simulations focus exclusively on galaxies with relatively high mass ($M_* \gtrsim 2 \times 10^8 \Msun$) compared to LAEs.
In order to understand the discrepancy between observations and simulations, follow-up spectroscopy to determine the accurate redshifts of galaxies in the quasar fields is important.
Moreover, the large scatter in the relationship between the SMBH and DMH masses due to their indirect connection through the host galaxy can cause the discrepancy.
The large scatter also appears in the simulation; \citet{Habouzit2019} reported that some SMBHs show much higher DMH mass than expected from the average black hole mass to DMH mass ratio.
Individual DMH mass measurements of quasars (e.g. \citealp{Fei2025}) also help investigate the relationship between the galaxy overdensity of quasars and their SMBH properties.

\subsubsection{Galaxy age} \label{subsubsec:age}
In Sec. \ref{subsubsec:host_galaxy} and \ref{subsubsec:smbh}, we assume that LAEs trace the underlying mass distribution.
However, different galaxy populations may show different distributions.
In general, galaxies in overdense regions are thought to form earlier and be older than their field counterparts, undergoing accelerated growth in their evolution (e.g. \citealp{Hatch2011, Koyama2013, Cooke2014}).
However, stacking analysis of LAE images at $z\sim6\mathchar`-7$ shows that they are typically young star-forming galaxies ($\sim1\mathchar`-3$ Myr; \citealp{Ono2010}).
Therefore, when discussing the overdense region of LAEs, one may need to be cautious about the LAEs' young age.
In fact, \citet{Ito2021} point out that the cross-correlation between LAEs and SFGs at lower-$z$ ($2<z<4.5$) is lower than SFGs' auto-correlations, and that LAEs are less abundant in SFG overdense regions. 
This suggests that LAEs may exhibit a formation epoch-dependent assembly bias and that they tend to be observed away from old populations.

Among our quasars, the spectrum of the host galaxy of J1512$+$4422 obtained by \textit{JWST} shows characteristics of a post-starburst galaxy \citep{Onoue2025}, and its mass-weighted age is $150_{-20}^{+20}$ Myr, which is much older than the typical LAE age.
Assuming that the galaxy conformity commonly observed in nearby galaxies \citep{Weinmann2006} --- that is, the high proportion of early-type galaxies in the vicinity of early-type central galaxies within a DMH --- also holds at high-z, even if no LAE overdensity is observed in the J1512$+$4422 field, it cannot be ruled out the possibility that there is an overdensity of galaxies too old to be observed as LAEs.
\citet{Kauffmann2013} suggest that this phenomenon occurs over the single halo scale, although the actual scale that manifests the galaxy conformity is still under debate \citep{Sin2017}. 
It would be valuable to measure the overdensity of the galaxy population other than LAEs in this field, as well as in the J0918$+$0139 and J1425$-$0015 fields, where LAE underdensity is observed, possibly resulting in similar situations.
It is possible that the number density of young galaxies (e.g. HAEs) is relatively small, while that of quiescent galaxies as old as the host galaxy of J1512$+$4422 is higher than that in blank fields.

\subsection{LAE overdensity around galaxies with a similar $M_*$ as our quasars}
\label{subsec:comparison_galaxy}
Our quasar samples leverage the advantage of known host galaxy $M_*$, allowing us to compare the LAE overdensity of our quasars with that of non-AGN galaxies with a similar $M_*$ to our quasars.
The stellar-to-halo mass relation anticipates that both the quasars and the non-AGN galaxies have similar DMH mass, resulting in a fair comparison of the LAE overdensities of them.

\subsubsection{LAE and galaxy data} \label{subsubsec:data}
Ideally, we would select LAEs around galaxies under conditions identical to those described in Sec. \ref{subsec:delta}. 
However, since no large LAE sample exists at our quasars' redshifts, we substitute a sample of LAEs at $5.65\leq z\leq 5.75$, which is close to this redshift, and LAE evolution during this interval ($\Delta z\sim0.48$ corresponds to $\sim0.1$ Gyr) is likely to be small.
\citet{Santos2016} present photometrically selected LAEs at $z\sim5.7$ in the survey area of 7 deg$^2$ in the following three survey fields: the Cosmic Evolution Survey (COSMOS; \citealp{Scoville2007}) field, the UKIDSS Ultra Deep Survey (UDS; \citealp{Lawrence2007}) field, and the SA22 field.
The effective area of the COSMOS field is 1.96 deg$^2$ (7056 arcmin$^2$).
They use multiband photometry ($BVgriJK$) in addition to NB816 and perform careful visual inspection to remove low-$z$ interlopers.
We use the LAEs in the COSMOS field out of those in the three fields.
As discussed below, the COSMOS field has abundant multi-wavelength data, allowing $M_*$ estimates for each galaxy via SED fitting.
They select LAEs with $\EW>25$~\AA\ from narrowband imaging, while our LAE sample is selected with $\EW>15$~\AA.
For a fair comparison, we reevaluate the LAE overdensity of our quasar fields using the LAEs with $\EW>25$~\AA, which can be selected by $z-NB872>0.9$ based on the simulation described in Sec. \ref{subsec:lae_selection}.

We select spectroscopically confirmed galaxies with $M_*$ equivalent to those of our quasars from the COSMOS Spectroscopic Redshift Compilation\footnote{\url{https://github.com/cosmosastro/speczcompilation}} \citep{Khostovan2026}.
The $M_*$ is evaluated by the SED fitting with \texttt{CIGALE}\footnote{\url{https://cigale.lam.fr/}} \citep{Boquien2019, Yang2020_XCIGALE}; for details on the parameters, please refer to \citet{Khostovan2026}.
There are 18 galaxies with $9.5\leq\log(M_*/\Msun)\leq11.5$ at $5.65\leq z\leq 5.75$, the same redshift range as the LAEs.
Taking the goodness-of-fit of the SED fitting and the reliability of the spectroscopic redshift, we select 5 galaxies that satisfy $\chi^2_\nu<5$ and $3\leq Q_f<9$ out of the 18 galaxies.
Here, $Q_f$ indicates the quality flag, and this criterion selects galaxies whose redshift confidence level is $\geq95\%$ and emission lines do not have broad line components
\footnote{In the case of $Q_f\geq10$, the spectrum shows broad line components, but we confirm that all of the 18 galaxies meet $Q_f<9$.}
.
We measure the LAE overdensity around these galaxies in basically the same manner as described in Sec. \ref{subsec:delta}, using 37 bright LAEs with $NB816<24.5$, which corresponds to the limiting magnitudes of NB872 measured by a $2\farcs0$ diameter aperture in our observations.
We evaluate $\langle\Sigma_\mathrm{LAE}\rangle$ as $37/7056=0.005$ arcmin$^{-2}$ in the COSMOS fields, where no quasars and no LAE protoclusters have been observed at $5.65\leq z \leq 5.75$.
We correct the effect of the edge regions using equation (\ref{eq:edge_correction}), but we do not correct the effect of masking low-quality regions.

\subsubsection{Do quasars affect the LAE overdensity?}
\begin{figure}
    \centering
    \includegraphics[width=\columnwidth, clip]{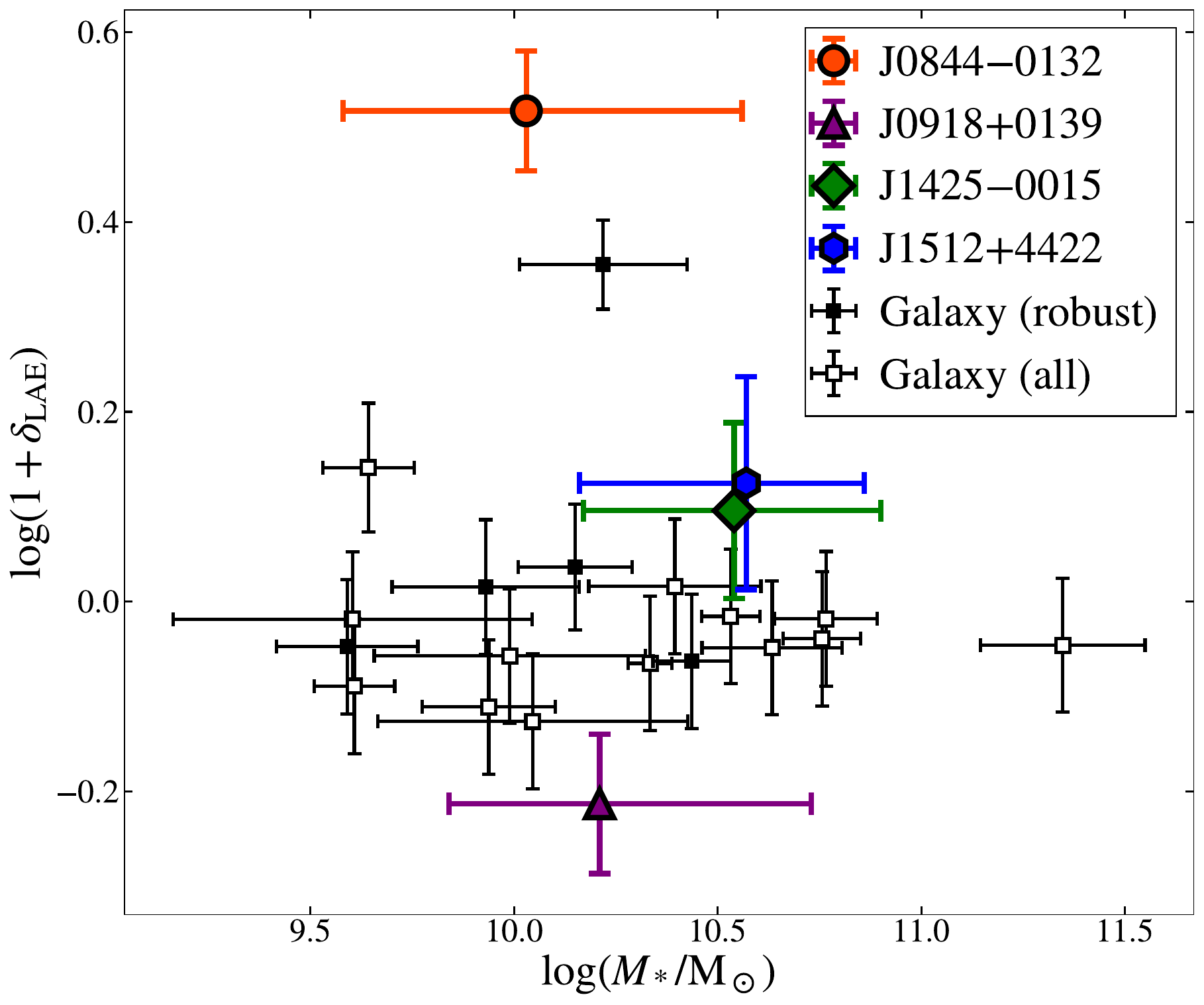}
    \caption{Comparison of surrounding LAE overdensities between our quasars and galaxies with similar $M_*$ in the COSMOS field with a uniform threshold of $\EW>25$~\AA\, selected to have $5.65<z<5.75$.
    The filled squares show the LAE overdensities of more robustly selected galaxies with reliable SED fitting and redshifts, while the open squares show those of all galaxies at $5.65\leq z\leq5.75$. 
    The LAE overdensities are calculated in the manner described in Sec. \ref{subsec:delta}.
    }
    \label{fig:lae_overdensity_comparison}
\end{figure}

Fig. \ref{fig:lae_overdensity_comparison} compares the LAE overdensities around our quasars with those of galaxies with a similar $M_*$.
We find that the LAE overdensity of J0844$-$0132 is much stronger than that of any of the galaxies with similar $M_*$, while the LAE overdensities in the remaining three quasars are comparable.
This suggests that the LAE overdensities in the three regions fall within the range predicted from their $M_*$, with only J0844$-$0132 potentially exhibiting a distinctive overdensity of $\delta_\mathrm{LAE}=2.29\pm0.48$.

The strong overdensity of J0844$-$0132 implies that some quasars could enhance the galaxy formation around them, an effect opposite to the negative feedback from quasars (e.g. photoevaporation).
\citet{Zana2022} use a cosmological simulation to demonstrate that quasar feedback can enhance the star formation activity.
They suggest that the outflow from the quasar supplies gas to its surroundings and, triggers shock waves that enhance star formation efficiency, making the quasar's vicinity a potential site for star formation.
The surrounding galaxies affected by this positive feedback from quasars may become more luminous, possibly resulting in the observed galaxy overdensity.
Positive feedback from quasars may also increase the intrinsic number of surrounding galaxies as they stimulate star formation in the surrounding gas.
However, it should be noted that the physical scale where positive feedback works efficiently is up to a few cMpc scale.
Otherwise, it is possible that J0844$-$0132 resides in a very massive DMH significantly exceeding the empirical mean stellar-to-halo mass ratio due to its large scatter (e.g. \citealp{Behroozi2019}).
To verify the scenario, radiation-hydrodynamical simulations with high spatial resolution and large box sizes are necessary to evaluate the effect of quasar feedback on their surrounding galaxies more accurately.
In addition, using ALMA to observe the outflows from quasars and investigating the galaxies (e.g. [\ion{C}{II}] emitters) around the quasars displaying the sign of outflows is important to observationally test the impact of quasar feedback.

\subsection{Other interpretations}

Since this study observes quasars at the last stage of the reionization epoch, LAE-deficient regions may correspond to neutral islands \citep{Giri2019, Keating2020, Nasir2020}, where the local IGM happens to have a high \ion{H}{I} fraction.
These regions can reduce the observable Ly$\alpha$ emission due to Ly$\alpha$ damping, which may explain why some quasars lie in apparently non-overdense regions.
However, it is unlikely that a quasar emitting intense radiation could coexist with these neutral islands.
Moreover, as shown in Fig. \ref{fig:ew_distribution} and mentioned in Sec. \ref{subsubsec:photoevaporation}, there is no tendency for $\EW$ to be small only close to the quasar.
Therefore, the influence of patchy reionization is implausible.

Recently, \citet{Fontanot2025} demonstrated that the distribution of surrounding galaxies around quasars can distinguish between triggering mechanisms contributing to the quasar ignition --- a galaxy merger event and a galactic disk instability --- using the \texttt{PLANCK MILLENNIUM simulation} and the semi-analytic model of GAlaxy Evolution and Assembly (\texttt{GAEA}; \citealp{DeLucia2024}).
Thus, two distinct mechanisms for quasar triggering also provide a plausible scenario to explain the diversity of the quasar environments.
In fact, J0844$-$0132 is the only quasar in our sample that has a companion galaxy with a separation of less than $0\farcs3$, which is indicative of a galaxy merger \citep{Ding2025}, although the separation is too close to be resolved in current observations.
Moreover, it is difficult to verify this scenario observationally since the triggering mechanisms for individual quasars are hard to distinguish from one another.
Further theoretical approaches (e.g. large cosmological simulations) are necessary to test the relation between the quasar environment and its triggering mechanisms.

Moreover, while high-$z$ quasars on average possess relatively high DMH masses of around $\sim10^{12}\,\Msun$ \citep{Arita2023, Eilers2024, Huang2026, Wang2026}, their DMH mass distribution might be broad, and the variety of DMH masses may be creating diverse environments.
Recently, \citet{Wang2026} investigate the diverse environments of 25 luminous quasars in the ASPIRE sample at $z>6.5$ to find that the distributions of line-of-sight velocities of [\ion{O}{III}] emitters around quasars in overdense regions are similar to those in massive DMHs with $\sim10^{13}\,\Msun$, while those around the other ASPIRE quasars are more consistent with those around DMHs with $5\times10^{11}\mathchar`-10^{12}\,\Msun$.
These results support the scenario that the DMH mass distribution is broad.
Further observations of many high-$z$ quasar fields with various host galaxy and SMBH properties are necessary to understand their intrinsic environments.
Even if the DMH mass remains the same, the assembly bias on galaxy distribution may be the driver to generate diversity.
This bias also depends on the observed galaxy population, making approaches that measure the environment with other galaxy populations equally important.
In this sense, surveys of other galaxy populations such as [\ion{O}{III}] emitters and H$\alpha$ emitters around many high-$z$ quasars are essential, which can be performed by \textit{JWST}/NIRCam narrowband imaging or slitless spectroscopy (e.g. \citealp{Wang2023}).
In addition, it would be valuable to search for dust-obscured galaxies or submillimeter galaxies that are invisible to rest-UV surveys to measure more accurate galaxy overdensities, using ALMA and Submillimetre Common-User Bolometer Array 2 on the James Clerk Maxwell Telescope.

Finally, increasing the sample size and expanding the survey area are important to overcome the cosmic variance.
As shown in Table \ref{tab:host_property}, $\langle\Sigma_\mathrm{LAE}\rangle$ of our quasar fields can vary significantly due to the cosmic variance, while the limiting magnitudes slightly differ among the quasar fields.
Therefore, it is also useful to perform narrowband imaging for more quasar fields with a larger effective area.

\section{Summary} \label{sec:summary}
This paper searches for LAEs surrounding four low-luminosity ($-24<M_{1450}<-22$) quasars at $z\sim6.18$ (J0844$-$0132, J0918$+$0139, J1425$-$0015, J1512$+$4422) with NB872 on Subaru/HSC to characterize their environments out to $\sim100$ cMpc.
These quasars are identified by the SHELLQs project and have been observed by \textit{JWST}, unveiling the properties of their host galaxies ($M_*$, SFR, sSFR) and SMBHs ($M_\mathrm{BH}$, bolometric luminosity, Eddington ratio).
We select LAE candidates with the colour criteria of equations (\ref{eq:limiting_magnitude})--(\ref{eq:nb_excess}) and perform a careful visual inspection.
We identified 88, 31, 28, and 23 LAEs in the fields of J0844$-$0132 (5173 arcmin$^2$), J0918$+$0139 (4837 arcmin$^2$), J1425$-$0015 (5092 arcmin$^2$), and J1512$+$4422 (5137 arcmin$^2$), respectively, not including the central quasars.
The surface number density of LAEs $>50$ cMpc away from the central quasars in our fields is almost comparable to that in blank fields at $z=5.7$ and $z=6.6$ \citep{Konno2018}.
We also compute the LAE overdensity using KDE with a Gaussian kernel, setting the bandwidth to 17.7 cMpc, the correlation length between galaxies and quasars at $z\sim6$ \citep{Arita2023}.
The main findings of this paper are summarized as follows:
\begin{enumerate}
    \item %%% Overdensity %%%
    The J0844$-$0132 field exhibits a high LAE overdensity with $\delta_\mathrm{LAE}=1.97\pm0.40$ at the quasar position.
    The LAE overdensities of the three other quasar fields are consistent with zero within their $1\sigma$ uncertainty.
    These results are summarized in Fig. \ref{fig:overdensity_map_corrected} and Table \ref{tab:host_property}.

    \item %%% EW distribution %%%
    We measure their $\EW$ based on the $z-NB872$ colour and test whether $\EW$ is correlated with the separation from the quasar ($\theta$).
    Using Kendall's $\tau$, we found no evidence that $\EW$ is depends on $\theta$ in all of our quasar fields.
    
    \item %%% photoevaporation %%%
    We measure the proximity zone sizes of our quasars from ground-based spectra.
    No LAEs are found within the proximity zones of our quasars.
    The absence of LAEs within the proximity zone of J0844$-$0132 is consistent with the expected number of LAEs based on $\Sigma_\mathrm{LAE}$ at the quasar position, making it difficult to verify whether the photoevaporation effect is important.
    In the remaining three fields, LAE overdensities are not observed over the proximity zone sizes.
    Hence, we conclude that the absence of LAE overdensities in the three fields is not due to the photoevaporation effect.
    Our fields show a broad range of environments from $\delta_\mathrm{LAE}=-0.11$ to $\delta_\mathrm{LAE}=1.97$, indicating that the LAE density around the quasar exhibits diversity independent of the photoevaporation effect.
    
    \item %%% young quasar %%%
    The proximity zone size of J0844$-$0132 is the smallest among our quasars ($R_\mathrm
    {p}=0.4449\pm0.0093$ pMpc), implying that it is a young quasar ($\lesssim10^4$ yr).
    If J0844$-$0132 is indeed young, it may suggest that even if photoevaporation works efficiently, it would not have had sufficient time to inhibit star formation in surrounding galaxies.

    \item %%% host/SMBH properties %%%
    None of the properties of the host galaxies ($M_*$, SFR, sSFR) or SMBHs ($M_\mathrm{BH}$, bolometric luminosity, Eddington ratio) are found to be strongly related to the LAE overdensity, contrary to expectations that galaxies with high $M_*$ or high $M_\mathrm{BH}$ reside in massive DMHs and thus should be detected in galaxy overdense regions.
    This lack of correlation might be due to the small size of our sample and the projection effect, or the large scatter between the DMH mass and the SMBH mass.
    
    \item %%% galaxy age %%%
    The mass-weighted age of the host galaxy of J1512+4422, which shows no significant LAE overdensity, is $150$ Myr \citep{Onoue2025}.
    The age is significantly older than the typical age of an LAE ($\sim1\mathchar`-3$ Myr).
    This suggests that even if a galaxy overdensity exists around the quasar, the surrounding galaxies may also be similarly old and show no significant Ly$\alpha$ emission.
    
    \item %%% QSO field vs non-QSO field %%%
    We find that the LAE overdensity in the J0844$-$0132 field is stronger than that of galaxies with similar $M_*$ at $z\sim6$, while the other quasar fields show a comparable LAE overdensity.
    Positive feedback from the quasar to promote the star formation activity of the surrounding galaxies may contribute to the large LAE overdensity in the field.
    However, the positive feedback works efficiently only up to a few cMpc scale; hence, it is also possible that the quasar resides in a very massive DMH, a few sigma above the empirical stellar-to-halo mass ratio.

    \item %%% other interpretation %%%
    The observed diversity of high-$z$ quasar environments may also be attributed to different triggering channels of the quasar activity, due to mergers and galactic disk instabilities, or the large scatter in the DMH mass distribution of quasars.
    
\end{enumerate}
This study demonstrates that the environments of high-$z$ quasars are diverse.
Nevertheless, it is highly intriguing that the DMH mass of high-$z$ quasars remains nearly constant over cosmic time \citep{Arita2023, Eilers2024, Huang2026, Wang2026}.
Furthermore, the observed diversity poses a significant challenge to many simulations that predict that high-$z$ quasars ubiquitously reside in high-density regions.
Even considering these factors, further observational constraints are necessary to fully understand the diverse quasar environments.
The LAE selection in this work uses imaging observations with the narrowband filter NB872.
Spectroscopic follow-up to determine the precise three-dimensional LAE distribution in the quasar fields would allow the overdensity to be better quantified.
The precise distribution will reveal a possible anisotropic quasar feedback and evaluate the accurate galaxy overdensity.
Therefore, future spectroscopic follow-up by, for example, Prime Focus Spectrograph (PFS; \citealp{Tamura2016}) will be useful to determine their robust redshifts. 
Observing more quasar fields with a narrowband filter will be useful to statistically evaluate the quasar impact on its large-scale environments, but not all quasars have a narrowband filter that corresponds to their redshifts.
To increase the sample size, slitless spectroscopy is suitable as it can identify galaxies around quasars without any target selection.
For example, the upcoming \textit{JWST}/NIRCam Wide Field Slitless Spectroscopy mode observation (GO 7519; PI: J. Arita) will search for [\ion{O}{iii}] emitters around 12 SHELLQs quasars, including the targeted quasars in this paper.
This observation will allow us to quantify the quasar environments, including the immediate vicinity of quasars, using general star-forming galaxies.
Although the FoV of NIRCam is much smaller than that of HSC, this observation enables us to directly compare the galaxy overdensity of quasars with that of normal galaxies with the same $M_*$ and redshift range.
These observational approaches will also enable us to verify the SMBH growth theory in the early Universe and inform more sophisticated cosmological simulations.

\section*{Acknowledgements}
We thank the referee, Jacklyn Champagne, for the insightful comments that greatly improved the manuscript.
We also thank Linhua Jiang, Roberto Gilli, and Fabio Vito for constructive discussions.
JA is supported by the Japan Society for the Promotion of Science (JSPS) KAKENHI grant number JP24KJ0858 and International Graduate Program for Excellence in Earth-Space Science (IGPEES), a World-leading Innovative Graduate Study (WINGS) Program, the University of Tokyo.
NK is supported by the Japan Society for the Promotion of Science through Grant-in-Aid for Scientific Research 21H04490, 25H00663, 25K01038, 25K01044.
YM is supported by the Japan Society for the Promotion of Science (JSPS) KAKENHI Grant No. 21H04494. 
MO is supported by the Japan Society for the Promotion of Science (JSPS) KAKENHI Grant Number 24K22894.
KI acknowledges support under the grant PID2022-136828NB-C44 provided by MCIN/AEI/10.13039/501100011033 / FEDER, UE.
SK is supported by the JSPS through Grant-in-Aid for Scientific Research 24KJ0058, 24K17101.

The Hyper Suprime-Cam (HSC) collaboration includes the astronomical communities of Japan and Taiwan, and Princeton University. The HSC instrumentation and software were developed by the National Astronomical Observatory of Japan (NAOJ), the Kavli Institute for the Physics and Mathematics of the Universe (Kavli IPMU), the University of Tokyo, the High Energy Accelerator Research Organization (KEK), the Academia Sinica Institute for Astronomy and Astrophysics in Taiwan (ASIAA), and Princeton University.  Funding was contributed by the FIRST program from the Japanese Cabinet Office, the Ministry of Education, Culture, Sports, Science and Technology (MEXT), the Japan Society for the Promotion of Science (JSPS), Japan Science and Technology Agency  (JST), the Toray Science  Foundation, NAOJ, Kavli IPMU, KEK, ASIAA, and Princeton University.

This paper is based [in part] on data collected at the Subaru Telescope and retrieved from the HSC data archive system, which is operated by Subaru Telescope and Astronomy Data Center (ADC) at NAOJ. Data analysis was in part carried out with the cooperation of Center for Computational Astrophysics (CfCA) at NAOJ.  We are honored and grateful for the opportunity of observing the Universe from Maunakea, which has the cultural, historical and natural significance in Hawaii.

This paper makes use of software developed for Vera C. Rubin Observatory. We thank the Rubin Observatory for making their code available as free software at \url{http://pipelines.lsst.io/}.

The Pan-STARRS1 Surveys (PS1) and the PS1 public science archive have been made possible through contributions by the Institute for Astronomy, the University of Hawaii, the Pan-STARRS Project Office, the Max Planck Society and its participating institutes, the Max Planck Institute for Astronomy, Heidelberg, and the Max Planck Institute for Extraterrestrial Physics, Garching, The Johns Hopkins University, Durham University, the University of Edinburgh, the Queen’s University Belfast, the Harvard-Smithsonian Center for Astrophysics, the Las Cumbres Observatory Global Telescope Network Incorporated, the National Central University of Taiwan, the Space Telescope Science Institute, the National Aeronautics and Space Administration under grant No. NNX08AR22G issued through the Planetary Science Division of the NASA Science Mission Directorate, the National Science Foundation grant No. AST-1238877, the University of Maryland, Eotvos Lorand University (ELTE), the Los Alamos National Laboratory, and the Gordon and Betty Moore Foundation.

%%%%%%%%%%%%%%%%%%%%%%%%%%%%%%%%%%%%%%%%%%%%%%%%%%
\section*{Data Availability}

The data in this paper will be shared upon reasonable requests to the corresponding author.



\bibliographystyle{mnras}
\bibliography{main} % if your bibtex file is called example.bib

%\begin{thebibliography}{99}
%\end{thebibliography}

%%%%%%%%%%%%%%%%%%%%%%%%%%%%%%%%%%%%%%%%%%%%%%%%%%

%%%%%%%%%%%%%%%%% APPENDICES %%%%%%%%%%%%%%%%%%%%%

\appendix

\section{Selection completeness of the colour selection criteria}
\label{apx:selection_completeness}

\begin{figure}
    \centering
    \includegraphics[width=\columnwidth]{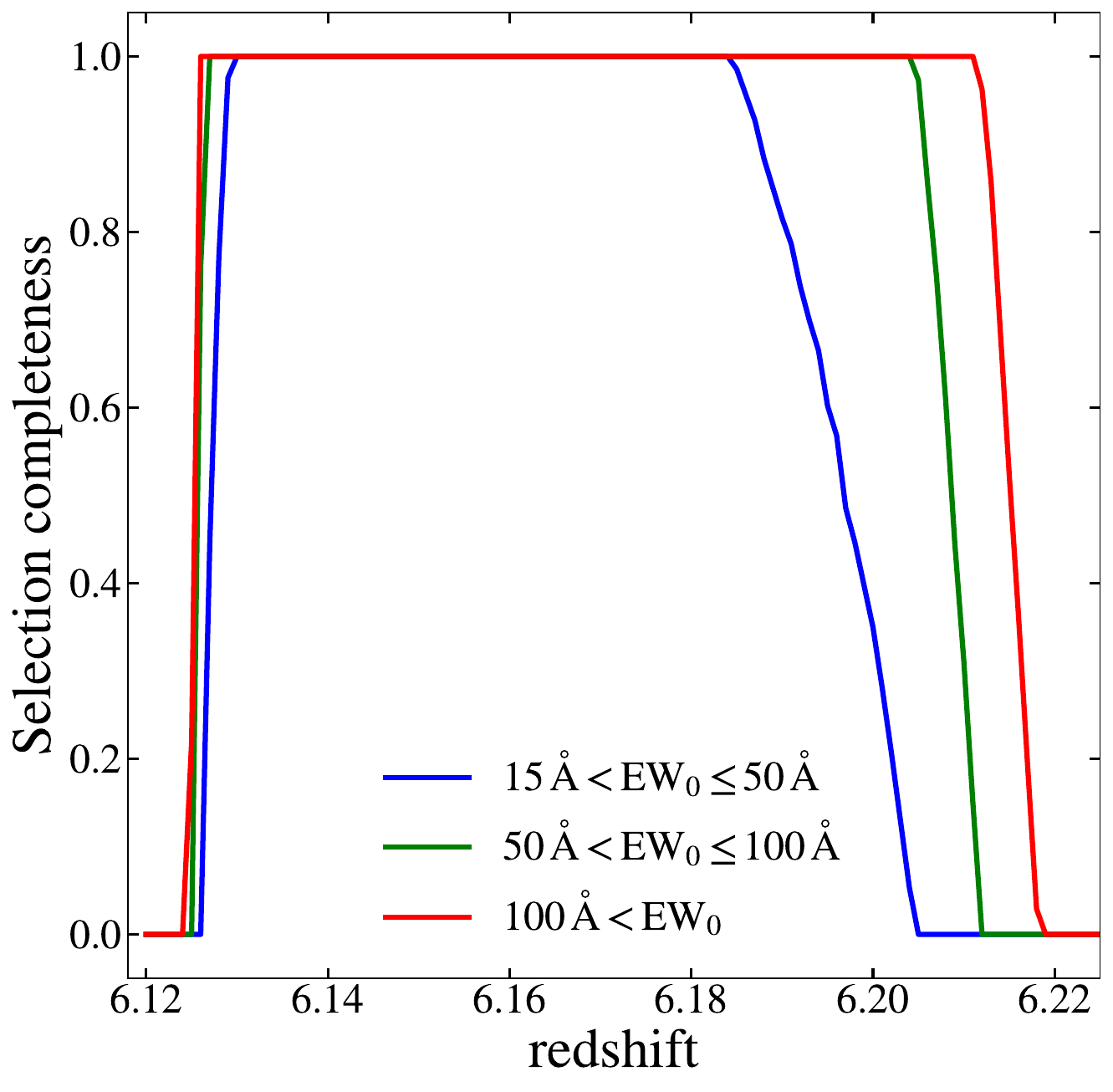}
    \caption{Selection completeness of our colour selection criteria as a function of redshift.
    The colour shows the $\EW$ ranges of LAEs.
    }
    \label{fig:selection_completeness}
\end{figure}

We compute the selection completeness of our selection criteria, which is defined as the recovery rate of LAEs within a given $\EW$ range that are selected by our colour selection.
We assume the intrinsic $\EW$ distribution that \citet{Shibuya2018} obtained at $z=5.7$: 
\begin{equation}
    \frac{\mathrm{d}N}{\mathrm{d}\EW}=N\exp\left(-\frac{\EW}{W_e}\right),
\end{equation}
where $W_e$ is the Ly$\alpha$ $\EW$ scale length of the exponential function.
We set $W_e$ to $138$~\AA\ from the results at $z=5.7$ \citep{Shibuya2018}.
Following the $\EW$ distribution, we generate 1000 mock LAEs with $\alpha=-2.5$ and FWHM of $250\,\mathrm{km\,s^{-1}}$ for each redshift bin between $z=6.12$ to $z=6.23$ with a step of $\Delta z=0.001$.
We compute the selection completeness of LAEs with three $\EW$ ranges: $15\,\mathrm{\text{\AA}}<\EW\leq50\,\mathrm{\text{\AA}}$, $50\,\mathrm{\text{\AA}}<\EW\leq100\,\mathrm{\text{\AA}}$, and $100\,\mathrm{\text{\AA}}<\EW$.
Their selection completeness is shown in Fig. \ref{fig:selection_completeness}.
We find that our selection criteria effectively select the LAEs at our quasars' redshifts.

\section{Relation between galaxy overdensity and the properties of the host galaxy and SMBH}
\label{apx:comparison_comprehensive}
\begin{figure}
    \centering
    \includegraphics[width=\columnwidth]{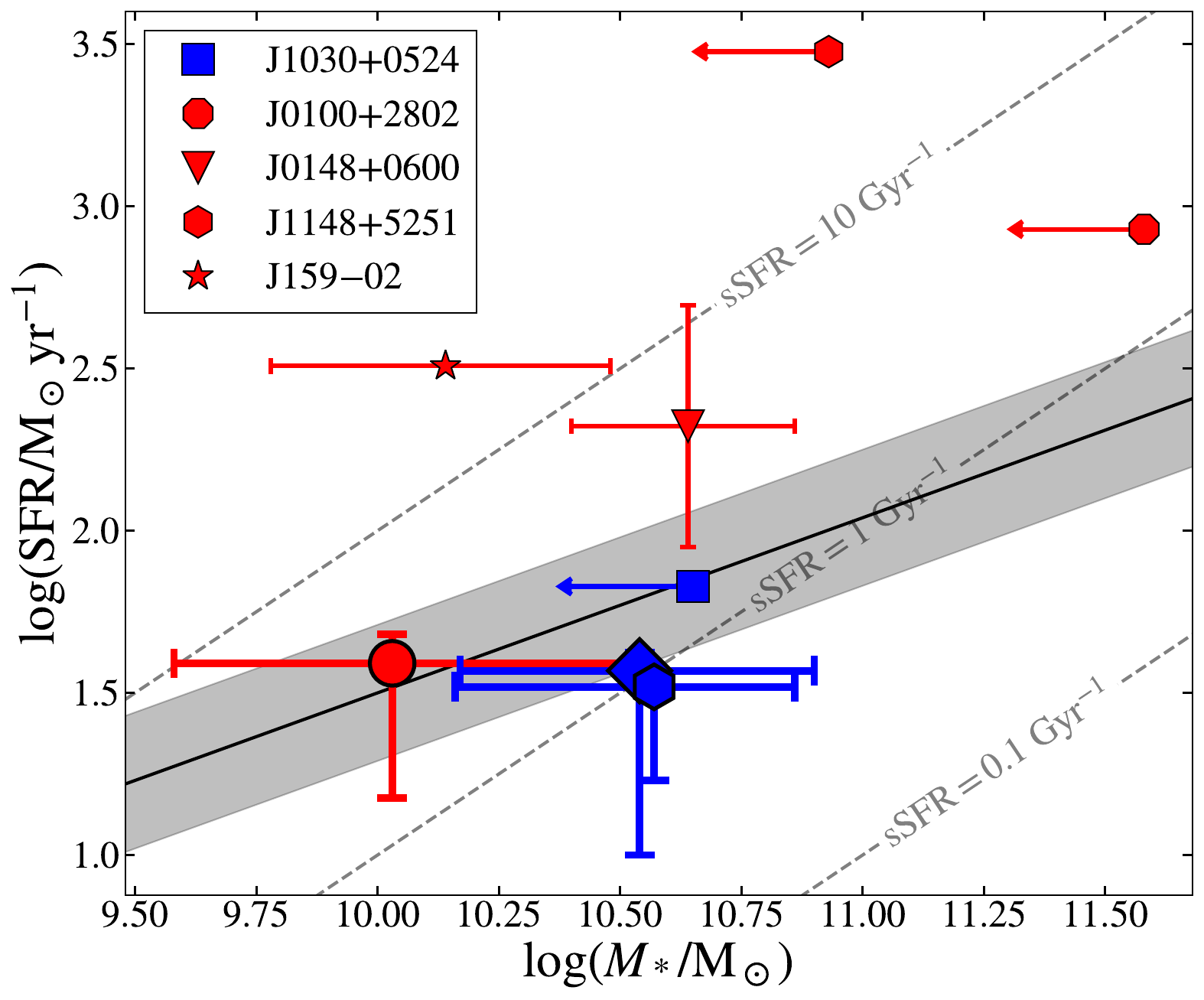}
    \caption{The relationship between $M_*$ and the SFR of the quasars.
    The plots for J0100$+$2802, J1148$+$5251, and J159$-$02 are added to Fig. \ref{fig:LAE_overdensity_host}.
    The red and blue symbols of the EIGER quasars show that the quasars reside in the overdense and non-overdense regions of [\ion{O}{III}] emitters, respectively.
    }
    \label{fig:mstar_sfr_appendix}
\end{figure}

\begin{figure}
    \centering
    \includegraphics[width=\columnwidth]{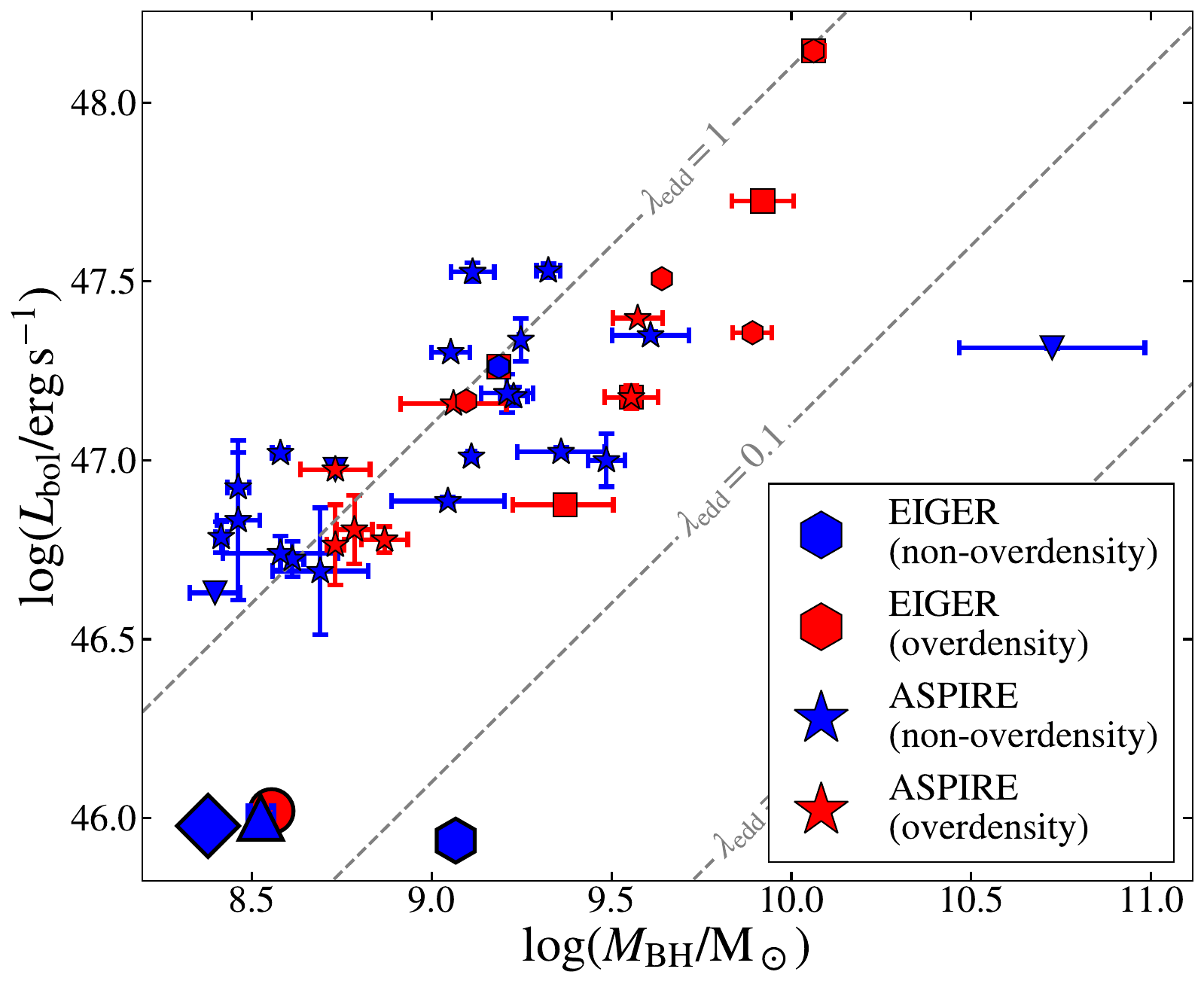}
    \caption{The SMBH masses and the bolometric luminosities of the quasars.
    The plots for the EIGER and ASPIRE quasars are added to Fig. \ref{fig:LAE_overdensity_smbh}.
    The red and blue symbols of the EIGER quasars show that the quasars reside in the overdense and non-overdense regions of [\ion{O}{III}] emitters, respectively.
    }
    \label{fig:mbh_lbol_appendix}
\end{figure}

In Sec. \ref{subsec:quasar_properties}, we investigate the relation between LAE overdensity and the properties of the host galaxy and SMBH.
Recently, the EIGER and ASPIRE projects have represented results on [\ion{O}{III}] emitter overdensity around high-$z$ quasars \citep{Eilers2024, Wang2026}.
Since the bias parameter could differ across galaxy populations and the spatial coverage for measuring overdensity varies between telescopes, a simple comparison between different analyses carries inherent risks.
Furthermore, the methods for measuring $M_*$ and SFR also differ between studies.
However, considering the limited number of samples available to investigate this correlation, we present the results in Fig. \ref{fig:mstar_sfr_appendix} and Fig. \ref{fig:mbh_lbol_appendix}, which include these [\ion{O}{III}] emitter samples
Readers should treat these results as purely indicative.

The host galaxies of four EIGER quasars have their $M_*$ and SFR measured: SDSS J010013.02$+$260225.8 (hereafter J0100$+$2802) at $z=6.33$; ULAS J014837.63$+$060020.0 (hereafter J0148$+$0600) at $z=5.98$; SDSS J114816.65$+$525150.2 (hereafter J1148$+$5251) at $z=6.42$; PSO J159.2257$-$02.5438 (hereafter J159$-$02) at $z=6.38$, whose redshifts are from \citet{Kashino2026}.
Their $M_*$ is measured by \citet{Yue2024} using the same method for J1030$+$0524.
The SFRs of J0100$+$2802, J0148$+$0600, J1148$+$5251, and J159$-$02 are evaluated in \citet{Wang2019a}, \citet{Li2020}, \citet{Maiolino2005}, and \citet{Decarli2018}, respectively.
To infer the SFRs, \citet{Maiolino2005} use the [\ion{C}{II}] line luminosity, while the other papers use FIR luminosity.
The overdensities of [\ion{O}{III}] emitters for the first three quasars and J1030$+$0524 are measured in \citet{Eilers2024}.
They show that the observed overdensities within a radius of 2 cMpc around J0100$+$2802, J0148$+$0600, J1148$+$5251, and J1030$+$0524 are $29\pm10,65\pm15,16\pm8$, and $3\pm4$, respectively.
This result suggests that J1030$+$0524 resides in a non-overdense region of [\ion{O}{III}] emitters while the other quasars reside in overdense regions.
While the overdensity around J159$-$02 is not evaluated in \citet{Eilers2024}, \citet{Kashino2026} show that the number of [\ion{O}{III}] emitters in the J159$-$02 field has a peak at the quasar redshift.
Therefore, we consider that J159$-$02 also resides in an overdense field of [\ion{O}{III}] emitters.
In addition, LBG overdensities around some of the quasars have also been measured.
\citet{Pudoka2024} show that the LBG overdensity of the J0100$+$2802 field is $\delta=4$ at $8.4\sigma$.
\citet{Morselli2014} evaluate the LBG overdensity of the J1148$+$5251 and J1030$+$0524 fields as $\delta=0.9$ and $2.0$ with significances of $\sigma_\delta=1.9$ and $3.3$, respectively.
These studies reveal that J0100$+$2802 and J1148$+$5251 reside in galaxy overdensities of both [\ion{O}{III}] emitters and LBGs, but that the J1030$+$0524 field does not show an overdensity of [\ion{O}{III}] emitters despite representing the LAE and LBG overdensities.

Fig. \ref{fig:mstar_sfr_appendix} shows the relationship between $M_*$ and the SFR of the quasars with that of our quasars.
Our quasars lie on the main sequence, whereas four quasars from the newly added EIGER quasars show starburst-like high SFRs above the main sequence, and all exhibit OIII emitter overdensities.
While this result is highly intriguing, it is difficult to immediately conclude a correlation between the overdensity and the SFR or sSFR of the quasar host galaxies. 
This is due to the heterogeneous sample consisting of LAEs and [\ion{O}{III}] emitters, the differences in volume used to measure the overdensity, the lack of uniformity in the SFR measurement methods, and the still limited data on the SFR and $M_*$ of the host galaxies.

Regarding the SMBH properties, all of the EIGER and ASPIRE quasars have their $M_\mathrm{BH}$ and bolometric luminosities measured from spectroscopic observations by \textit{JWST}.
We refer to the measurements in \citet{Yue2024} and \citet{Wang2026} and the overdensity of [\ion{O}{III}] emitters in \citet{Eilers2024} and \citet{Wang2026} for the EIGER and ASPIRE quasars, respectively.
Among the EIGER and ASPIRE quasars, one and seven quasars out of five and 25 quasars reside in the overdense regions of [\ion{O}{III}] emitters.

Fig. \ref{fig:mbh_lbol_appendix} shows the relationship between $M_\mathrm{BH}$ and the bolometric luminosity of the EIERG and ASPIRE quasars in addition to those in Fig. \ref{fig:LAE_overdensity_smbh}.
Even after adding the EIGER and ASPIRE quasars to those in Sec. \ref{subsubsec:smbh}, no correlations between galaxy overdensity and SMBH properties are confirmed.

%%%%%%%%%%%%%%%%%%%%%%%%%%%%%%%%%%%%%%%%%%%%%%%%%%

% Don't change these lines
\bsp	% typesetting comment
\label{lastpage}
\end{document}